\documentclass{aa}
\usepackage[varg]{txfonts}
\usepackage{natbib,twoopt}
\usepackage[bookmarks=false]{hyperref}
\usepackage{graphicx}
\usepackage{float}
\usepackage{color}
\usepackage{verbatim}
\usepackage[normalem]{ulem}

\begin{document}
\title{The stellar content of the XMM-Newton slew survey}
\author{S. Freund \and J. Robrade \and C. Schneider \and J.H.M.M. Schmitt}
\institute{Hamburger Sternwarte, Universit\"at Hamburg, 21029 Hamburg, Germany \\
e-mail: sebastian.freund@uni-hamburg.de; jrobrade@hs.uni-hamburg.de; cschneider@hs.uni-hamburg.de; jschmitt@hs.uni-hamburg.de}
\abstract{}
{We present a detailed analysis of the stellar content of the current version of the XMM-Newton slew survey (XMMSL2).}
{Since stars emit only a small fraction of their total luminosity in the X-ray band, the stellar XMMSL2 sources ought to have relatively bright optical counterparts. Therefore the stellar identifications are obtained by an automatic crossmatch of the XMMSL2 catalog with the first Gaia data release (Gaia DR1), the 2MASS and the Tycho2 catalogs. The reliability of this procedure is verified by a comparison with the individually classified 
\textit{Einstein Observatory} medium sensitivity survey X-ray sources and by a crossmatch with the Chandra Source Catalog.}
{We identify 6815 of the 23\,252 unique XMMSL2 sources to be stellar sources, while 893 sources are flagged as unreliable. For every counterpart a matching probability is estimated based upon the distance between the XMMSL2 source and the counterpart. Given this matching probability the sample is expected to be reliable to 96.7\,\% and complete to 96.3\,\%. The sample contains stars of all spectral types and luminosity classes, with late-type dwarfs having the largest share. For many stellar sources the fractional contribution of the X-ray band to the total energy output is found above the saturation limit of previous studies ($L_\mathrm{X}/L_\mathrm{bol}=10^{-3}$), because the XMMSL2 sources are more affected by flares due to their short exposure times of typically $6$~s. A comparison with the ''Second ROSAT all-sky survey (2RXS) source catalog'' shows that about 25\,\% of the stellar XMMSL2 sources are previously unknown X-ray sources. The results of our identification procedure can be accessed via VizieR.}
{}
\maketitle

\section{Introduction}
\label{sec: introduction}
\citet{cat75} were the first to detect coronal X-ray emission from a star
other than the Sun in
the bright active binary system Capella, albeit at a level much 
brighter than typical solar X-ray emission levels.  
Later, X-ray observations with the {\it Einstein Observatory} \citep{vai81} 
and then with {\it ROSAT} showed X-ray emission to be ubiquitous for 
almost all types of stars \citep{schmitt95,schmitt97,huensch98a,huensch98b}.

The X-ray properties of stars are usually investigated either by pointed observations of selected X-ray sources, e.g., nowadays
with the {\it XMM-Newton} \citep{XMM-mission} or with the {\it Chandra} 
satellite, or by all-sky surveys, e.g., the ROSAT all-sky survey \citep{RASS-catalog}
or, in the future, with the eROSITA all-sky survey \citep{erosita};
such all-sky surveys have the advantage of delivering large samples of X-ray sources that are not biased by the selection of specific sources or specific sky regions. 
The same applies to the XMM-Newton slew survey \citep[XMMSL;][]{sax08}, which delivers data
in a spectral range identical to the upcoming eROSITA survey.  The XMMSL, however, is
somewhat special in the context of X-ray surveys:
The XMM-Newton satellite also collects X-ray data while
slewing from one pointed observation to the next and these data 
form the basis of the XMMSL, 
which is regularly updated with the mission.   

Naturally, in contrast to ''true'' all-sky survey such a survey is quite inhomogeneous, but the XMMSL covers -- in its current version (XMMSL2) -- 
already 84\,\% of the sky and includes 29\,393 detections of 23\,252 unique X-ray sources.  
We are specifically interested in the stellar content
of the XMMSL, hence our task at hand is the development of a procedure
that distinguishes stellar sources in the XMMSL from other classes of X-ray emitters such as galaxy clusters, active galactic nuclei (AGN) and others
as reliably and completely as possible. 
Due to the large number of XMMSL2 sources, this
identification process can obviously not be carried out individually ''by hand'', rather an automatic method is required that utilizes the known properties
of stellar (coronal) X-ray sources.

Stellar X-ray sources are relatively ''faint'', when measured 
in terms of the fractional contribution of the X-ray band to the 
total energy output, i.e., the $L_\mathrm{X}/L_\mathrm{bol}$-ratio. 
For example, early-type stars typically satisfy $L_\mathrm{X}/L_\mathrm{bol} \approx 10^{-7}$ \citep{pal81,ber97}, and their X-ray emission is generated through radiative instabilities in their radiatively driven stellar winds.
In contrast, the X-ray emission observed from
late-type, ''cool'' stars is produced in hot coronae, and 
magnetic fields are thought to play a fundamental role for the
coronal physics of stars \citep{pev03}. 
The observed X-ray luminosities of late-type stars vary enormously, both in
individual cases as well as in a sample of stars. In the case of 
flares the X-ray flux can increase by orders of magnitude over time scales of minutes to hours.  Also, late-type stars may show 
modulated X-ray emission on time scales of years related to activity cycles  \citep{Hempelmann_2003,Favata_2008,Ayres_2009,Robrade_2012} similar to 
the solar cycle, in addition the X-ray flux of a given star may vary on the
time scale of rotation on typically a time scale of a few days and possibly
longer.  As a class, late-type dwarfs show a rather well defined 
maximum fractional X-ray emission of about $L_\mathrm{X}/L_\mathrm{bol} = 10^{-3}$ during 
so-called quasi-quiescent periods, i.e., during times without obvious strong flares \citep{vil84, agr86, fle88, pal90}.  A similarly well defined lower limit does not exist, but \citet{schmitt97} showed the existence of
a minimum X-ray surface flux of about $10^4\,$erg\,s$^{-1}$\,cm$^{-2}$ for
dwarf stars, which results in a
$L_\mathrm{X}/L_\mathrm{bol} \approx 10^{-7} - 10^{-6}$ for solar analogs.

Stars off the main-sequence also show X-ray emission, and 
the X-ray luminosity can be very high, especially for giants that are part of a binary system, e.g., RS CVn systems and related systems \citep{wal78, dem93}. However, little to no X-ray emission is found for red giants beyond the so-called dividing line \citep{linsky79,haisch91,huensch96}.

The low fractional X-ray luminosity stellar X-ray sources implies that 
counter parts of these sources ought to be relatively bright in the optical band.  
Hence, any star, detected for example in the XMMSL, will also
be detected in an optical survey of sufficient sensitivity.  In this context
the currently operating Gaia mission \citep{Gaia-mission} is particularly
relevant, since Gaia will eventually produce a complete optical catalog
down to a magnitude of 20 as well as parallaxes, which allows to easily
distinguish nearby stellar sources from more distant galactic and extragalactic
sources.  In November 2016 the first data release of the Gaia optical all-sky survey was issued by the \citet[][Gaia DR1]{GaiaDR1}, and we can therefore
start to tap the Gaia potential in our effort to 
identify the stellar XMMSL sources by a crossmatch with the Gaia DR1.

The plan of our paper is then as follows. In Sect.~\ref{sec: observations} we present the properties of the XMMSL and the Gaia DR1 catalog
as well as the complimentary catalogs used in this paper. 
In Sect.~\ref{sec: matching procedure} we describe our matching procedure and estimate the expected completeness and reliability of our stellar identification based upon the matching probability of the individual counterparts. We present our results and compare our identifications with those of \citet{sax08} in Sect.~\ref{sec: results}. Additionally, we test the reliability of our automatic matching procedure by applying it to the Extended Medium-Sensitive Survey (EMSS) \citep{gio90,sto91} of the {\it Einstein Observatory}, whose
sources have been individually classified by spectroscopy, and by performing a crossmatch with the Chandra Source Catalog. In Sect.~\ref{sec: RASS counterparts} we compare the X-ray fluxes of
the stellar XMM-Newton slew survey sources with the corresponding fluxes measured in ROSAT all-sky survey \citep{RASS-catalog}. 
The properties of the stellar sample of the XMMSL2 sources are presented in Sect.~\ref{sec: properties} and in 
Sect.~\ref{sec: Conclusions} we draw our conclusions.

\section{Catalog suite}
\label{sec: observations}

We first provide short descriptions of the various catalogs used in this paper.

\subsection{XMMSL catalog}
\label{sec: XMMSL2 catalog}
For our identification, we use the ``clean'' version of the XMMSL2 catalog as 
X-ray input catalog, which we refer to as XMMSL2 catalog hereafter; for a
detailed description of the catalog and its creation we refer to
\citet{sax08}, who describe all methods of the production of the first XMM-Newton slew survey catalog in detail, which are very similar to those of the XMMSL2 catalog.  Briefly, 
this catalog contains detections with a detection likelihood of $DET\_ML > 10.5$ in general and of $DET\_ML > 15.5$ for sources with higher than the usual 
background. 
The positional accuracy is typically about $8$~arcsec and -- 
according to \citet{sax08} -- about 4\,\% of the sources detected in the 
full band (as well as 0.7\,\% and 9\,\% of the sources detected in the soft and hard band, respectively) are spurious. 

All XMM-Newton slews are treated individually during the creation of the XMMSL2 
catalog, so that every X-ray detection leads to a new entry in the catalog, even if the same source has been detected in a previous slew. 
In a second step, detections lying within $30$~arcsec in different slews are then considered to be 
multiple detections of the same source, and are given the same source name. 
Therefore, the 29\,393 XMMSL2 detections actually come from 23\,252 unique X-ray sources. 
For our identification we only use the unique XMMSL2 sources and coordinates of the detection with the highest detection likelihood. Further, we use the median 
X-ray flux for sources with multiple detections.

In the XMMSL2 catalog the source count rates are given in three different energy bands,
i.e., the total band ($0.2-12$~keV), the soft band ($0.2-2$~keV), and the hard band ($2-12$~keV). 
%Coronal X-ray sources are typically soft X-ray emitters, unless they are heavily absorbed, and we require that they are detected in the soft band or the total energy band. 

\subsection{Gaia DR1}
\label{Gaia DR1}

The Gaia DR1 catalog contains the positions and $G$ band magnitudes of 1.1 billion sources. 
For a subset of 2 million stars, parallaxes and proper motions have been 
calculated from information provided by the HIPPARCOS and Tycho2 catalogs;
this subset is called the Tycho-Gaia astrometric solution (TGAS). 
The positional and photometric uncertainties of all catalog entries are negligible compared to the uncertainties in the X-ray data
(better than $10$\,mas and $0.03\,$mag, respectively).

Unfortunately, the Gaia DR1 catalog has a only preliminary character 
\citep{GaiaDR1}, and for our purpose, the relevant known limitations are its incompleteness for very bright sources $\lesssim 7$~mag, for sources with high proper motion, for extremely blue or 
red sources, and for sources located in dense areas on the sky and or in
binary systems.  To overcome these limitations 
and to obtain color information for spectral type classification, we consider complementary catalogs.

\subsection{Complementary catalogs}
\label{sec: complimentary catalogs}
To obtain colors and hence
spectral types for the X-ray counterparts, we use the 2MASS catalog \citep{2MASS}, i.e., an infrared catalog
that is particularly useful for stars of late spectral type (cf. Fig.~\ref{fig: minimalbrightness}),
and the Tycho2 catalog \citep{Tycho2} to identify the brighter sources.
In addition, we consider information provided by the Bright Star catalog \citep{BrightStar} and the catalog by \citet{Lepine} of bright M dwarfs (Lepine catalog). In the following we describe the matching procedure only for the Gaia DR1, 2MASS and Tycho2 catalog that provide the vast majority of our stellar identifications, but the presented method is also applied to the BrightStar and the Lepine catalog and appropriate matching distances and probabilities are estimated for these catalogs.

We expect essentially all stellar XMMSL2 sources to have a 2MASS counterpart. Since the completeness of the 2MASS catalog is $>99\,\%$, our procedure is not influenced by the incompleteness of the catalogs used for the identification.

\section{Data analysis \label{sec: matching procedure}}
Our matching procedure is based upon the angular distance between the XMMSL2 X-ray sources and potential stellar catalog counterparts (where we correct the position of the stellar counterpart for proper motion, if the proper motion is given in the catalog). It includes optical/NIR brightness cuts to limit the number of chance alignments with faint sources that are unlikely responsible for the X-ray emission. The appropriate magnitude cuts depend on the 
X-ray detection limit, i.e. the conversion between observed count rate and flux, as well as the expected ratio between X-ray and optical/NIR fluxes.
We then test our procedure against random sources, motivate the used matching distances and discuss the further applied selection procedures;
a flowchart of our matching procedure is given in Appendix~\ref{sec: flowchart}.

\subsection{X-ray fluxes \label{sect: fluxes}}
The conversion between the measured count rates and the derived X-ray fluxes depends on the spectral model assumed for the X-ray source. Therefore, we do not adopt the X-ray fluxes given in the XMMSL2 catalog that are estimated by applying a spectral model typical for AGN, but we use our own conversion adopting a spectral model, appropriate for stellar X-ray sources.
%In order to estimate the minimal optical brightness of a stellar source in the XMMSL2, we convert the measured count rates into X-ray fluxes using a spectral model appropriate for stellar X-ray sources.
Specifically, we assume optically thin emission and
adopt an APEC thermal plasma model with a temperature of $5\times 10^6$~K and solar metallicity.
We neglect interstellar extinction, because we expect to find most of the stellar counterparts within 150~pc. However, a few sources located in star forming regions might be affected by the interstellar absorption, but with the data at hand we cannot identify these sources.
Furthermore, we convert these fluxes into the ROSAT band to compare the XMMSL2 X-ray fluxes with previous measurements and
use a count rate [cts\,s$^{-1}$] to flux conversion factor of $1.24\times 10^{-12}$~erg\,cm$^{-2}$\,s$^{-1}$
for the soft and total band. The conversion factor 
is relatively insensitive to the assumed temperature; for the $2 - 35\times 10^6$~K range,
it changes by 6 \% and 15 \% for the total and soft band, respectively. 
The hard band is generally less suitable to observe stellar X-ray sources, because coronal X-ray sources are typically rather soft X-ray emitters, unless they are heavily absorbed, and the effective area of the XMM-Newton decreases for high energy photons. We formally adopt a flux conversion factor of $5.93\times 10^{-10}~\mathrm{erg\,cm^{-2}\,s^{-1}}$ per count rate of $1~\mathrm{cts\,s^{-1}}$ for the hard band, but this value is -- naturally -- very sensitive to the assumed model temperature, and we do not expect stellar sources to be detected only in the hard band.

The detection limit of the XMMSL2 catalog is typically $\sim~0.4$~cts\,s$^{-1}$  for a source passing through 
the center of the detector at a typical background level. With the adopted
conversion factor this corresponds to an X-ray flux of $5\times 10^{-13}$~erg\,cm$^{-2}$\,s$^{-1}$, which we use to derive the minimum optical brightness of potential stellar counterparts.

\subsection{Minimal optical brightness of the stellar XMMSL2 sources and magnitude cutoff}
\label{sec: minimal brightness}

\begin{figure}[t]
%\captionsetup{labelfont=bf}
\resizebox{\hsize}{!}{\includegraphics{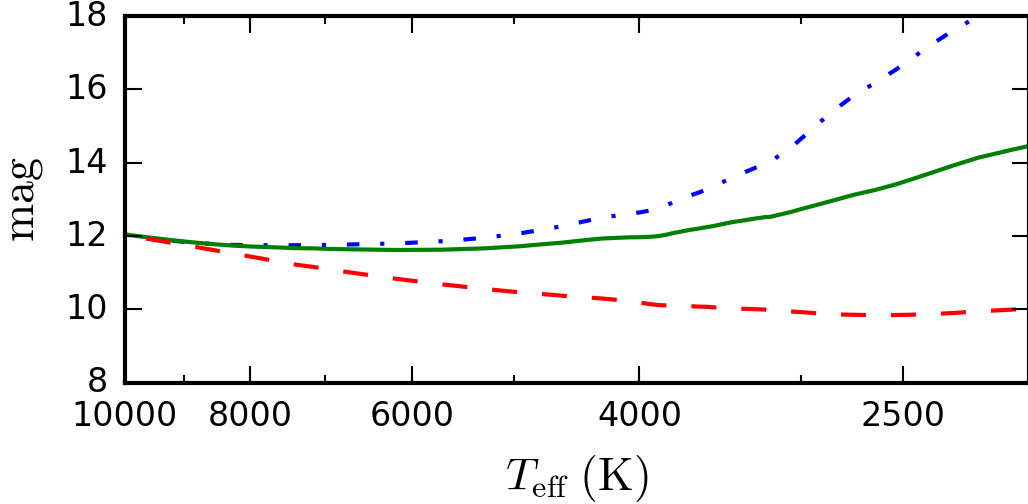}}
\caption{Apparent magnitude in different photometric bands for a 
star with saturated X-ray emission at the detection limit of the XMMSL2 catalog vs.
effective temperature. Blue dash-dotted line: V band; green solid line: G band; red dashed line: J band.}
\label{fig: minimalbrightness}
\end{figure}

With our estimate of the limiting XMMSL2 flux and the saturation limit of stellar X-ray 
emission, we can compute the minimal  bolometric flux of a possible
stellar counterpart to an XMMSL2 X-ray source. Given this minimal bolometric flux,
we use Table~3 of \citet{color-table} to calculate the minimum optical brightness in different photometric bands as a function of the effective temperature (cf. Sect.~\ref{sec: Estimation of additional stellar properties}), again neglecting interstellar absorption, and show the computed
magnitudes in the $V$, $G$ and $J$ bands vs. effective temperature
in Fig.~\ref{fig: minimalbrightness}.  As is clear from Fig.~\ref{fig: minimalbrightness},
all stellar XMMSL2 sources ought to be relatively bright in the optical with $G \lesssim 14$~mag.  Furthermore, the stellar sources are quite bright also in the infrared 
band with $J \lesssim12$~mag; in particular, very late spectral type dwarfs can be quite faint in the $V$ and $G$ bands, but should still be bright in the $J$ band of 2MASS. 

In practice, the magnitude limit shown in Fig.~\ref{fig: minimalbrightness} is 
not sharp.
First of all, there appears to be some intrinsic scatter in the saturation limit \citep{pizzo03} and, second, sources might be caught during an X-ray flare
during the slew survey observations and therefore produce more
X-ray flux than ''expected''.
Hence, the $L_\mathrm{X}/L_\mathrm{bol}$ ratio of coronal sources detected in 
the XMMSL2 could be higher than the saturation limit and their optical brightness lower than the predictions shown in Fig.~\ref{fig: minimalbrightness}. 
For example, \citet{stelzer06} report a flare with a peak X-ray luminosity 200-300 times above the quiescence emission and with an increase in optical brightness 
of $\Delta V=6$~mag for the star \mbox{LP 412-31}; however, these extreme flare events are 
quite rare and it is unlikely that XMM-Newton slews over a star during the 
peak of such an extreme flare. 

Yet to allow for some margin in these cases, we adopt a magnitude cutoff at 
$G=16$~mag for Gaia sources, at $J=12$~mag for 2MASS sources and no cutoff for Tycho2 sources; however, if neither a Tycho2 nor a 2MASS counterpart is found, 
the Gaia cutoff is set to $G=15$~mag.
These cutoff values are clearly sufficient to find all potential stellar counterparts 
emitting at the X-ray saturation level. The achieved stellar activity margin depends on spectral type, e.g. for stars with an effective temperature of $3000$~K (spT: M5V) the X-ray flux limit is about $\log(F_\mathrm{X}/F_\mathrm{bol}) < -2.2$ at minimum optical brightness.
We remark in passing that the
adopted magnitude cutoffs are well above the completeness limits of the Gaia and 2MASS catalog, so that all stellar XMMSL2 sources ought to be included in the catalogs.

\subsection{Random matches and magnitude distribution}
\label{sec: magnitude cutoff and random matches}

With its 1.1 billion sources the mean distance between two Gaia entries is about
$20~\mathrm{arcsec}$, thus finding a Gaia catalog entry in the vicinity of an XMMSL2 source
is not surprising.  In order to investigate the influence of random coincidences 
on our matching procedures, we carry out Gaia identifications with 
randomly generated X-ray sources. 
Since both the XMMSL2 catalog and the stellar catalogs chosen for matching
are nonuniform, it is important to preserve the global spatial distribution 
of the X-ray sources in the randomly generated X-ray samples; this
is achieved by using all cataloged XMMSL2 sources but shifting their
positions uniformly between a distance of
$240~\mathrm{arcsec}$ and $1200~\mathrm{arcsec}$ along a randomly chosen direction. 

\begin{figure}[t]
%\captionsetup{labelfont=bf}
\resizebox{\hsize}{!}{\includegraphics{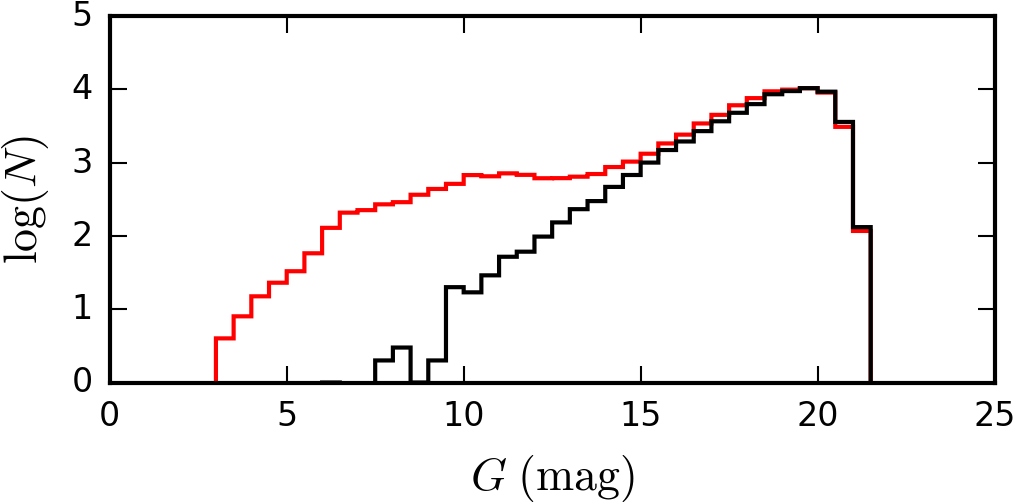}}
\caption{Magnitude distribution of the Gaia counterparts. Red line: Magnitude
distribution of 
the Gaia associations of the real XMMSL2 sources; black line: Magnitude
distribution of the Gaia associations with the randomly generated sources}
\label{fig: mag dist Gaia}
\end{figure}

In Fig.~\ref{fig: mag dist Gaia} we show the thus obtained
$G$~magnitude distribution of all Gaia matches (using
a matching distance of $20~\mathrm{arcsec}$ to the XMMSL2 sources) 
as well as that to the randomly generated 
XMMSL2 sources. Obviously, these distributions substantially differ from
each other. While the Gaia magnitude distribution of the true XMMSL2 sources
is bimodal with a broad first peak near $ G \approx 10$~mag, the Gaia magnitude distribution 
of the random XMMSL2 sources steadily increases up to the 
magnitude cutoff of the Gaia catalog near $G \approx$ 20~mag. 
Interestingly, the identified number of true XMMSL2 sources exceeds that of
randomly generated XMMSL2 sources up to $G \approx$ 20~mag. Thus,
even at faint magnitudes some of the Gaia counterparts appear to be the correct, 
albeit not necessarily stellar, 
identifications. However, it is also clear that for a magnitude $G \approx 14$~mag, the chance to obtain a random match exceeds 50\,\% using the matching distance of $20~\mathrm{arcsec}$.

\subsection{Choice of the matching distance}
\label{sec: matching distance}

\begin{figure}[t]
%\captionsetup{labelfont=bf}
\resizebox{\hsize}{!}{\includegraphics{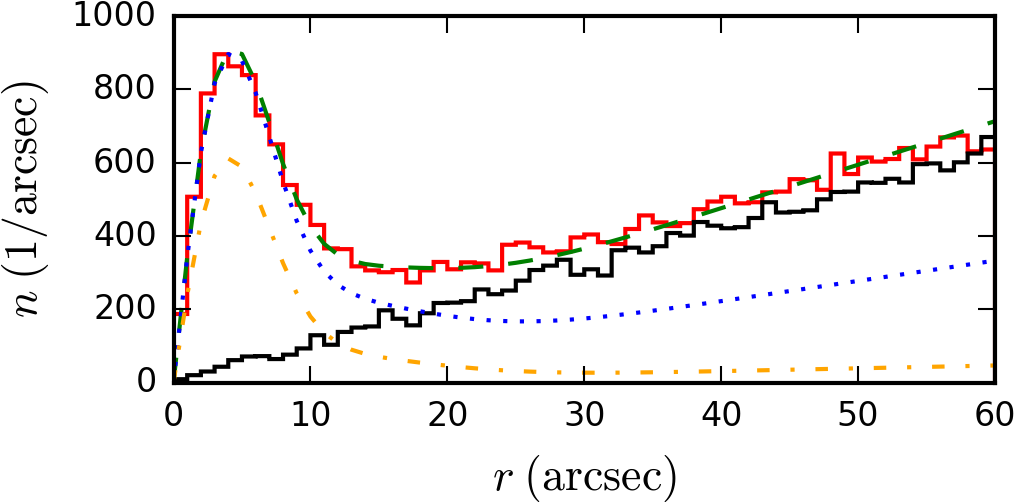}}
\caption{Distribution of the distances between the XMMSL2 sources and the stellar counterparts:
Red solid line: Histogram of the distances between the XMMSL2 sources and the Gaia counterparts, black solid line: Histogram of the distances between the randomly generated sources and the Gaia counterparts. The dashed and dotted curves represent the best fits of Eq.~\ref{equ: model distance} for the different catalogs with the parameters of Table~\ref{tab: fitted parameters}. Green dashed curve: Gaia catalog, blue dotted curve: 2MASS catalog, orange dash-dotted curve: Tycho2 catalog}
\label{fig: dist_plot}
\end{figure}

Next, we consider the (differential) number of matched XMMSL2-Gaia sources (choosing only Gaia entries 
with $G<15$~mag) as a function of matching distance and 
show the resulting histograms for the real XMMSL2 sources and the randomly generated XMMSL2 sources in Fig.~\ref{fig: dist_plot}. Again, the two distributions
differ substantially: The distribution of the randomly generated sources increases linearly as expected, while the distribution of the real XMMSL2 sources is bimodal. At small matching distances it is dominated by a Gaussian-type distribution up to a distance of $15$~arcsec. We find this distribution to be better fitted by a double Gaussian than by a single Gaussian distribution for the XMMSL2 sources. However, this is only an empirical description without any deeper physical meaning. At larger distances the distribution of the real XMMSL2 sources approximates the linear distribution of the randomly generated sources. The peak at small distances contains the true matches, while
the linearly increasing population of matches represents random associations.
In Fig.~\ref{fig: dist_plot} we also plot the corresponding distributions
resulting from the 2MASS catalog using a magnitude limit of $J=12$~mag and the full Tycho2 catalog. These distributions are qualitatively 
similar to the Gaia distribution
and differ only quantitatively because of the smaller number of catalog entries.

\subsubsection{Matching probability} 
\label{sec: matching probability}

The distribution of the real XMMSL2 sources can be well fitted with a double Gaussian (describing the uncertainty in the XMMSL2 positions) and a linear curve (describing the random matches) using the ansatz
\begin{equation}
n(r) = A\cdot r \cdot \exp \left(-\frac{r^2}{2\sigma_{1}^2}\right) + B\cdot r \cdot \exp \left(-\frac{r^2}{2\sigma_{2}^2}\right) + M\cdot r.
\label{equ: model distance}
\end{equation}
In Eq.~\ref{equ: model distance} the parameters $\sigma_1$ and $\sigma_2$ are the standard 
deviations of the Gaussian distributions, which are independent of the matching catalog; we find $\sigma_1 = 4.0$~arcsec and $\sigma_2 = 9.9$~arcsec. The values for the
parameters $A$, $B$, and $M$ depend on the source densities of the catalogs.  In 
Table~\ref{tab: fitted parameters} we provide the best fit parameters for the catalogs used in this paper, and in Fig.~\ref{fig: dist_plot} we give
a visual representation of the best fit curves.

\begin{table}[t]
\centering
\caption{Fitted parameters of Eq.~\ref{equ: model distance}}
\label{tab: fitted parameters}
\begin{tabular}{llll}
\hline \hline
 & Gaia & 2MASS & Tycho2 \\
\hline
$A$ [1/arcsec$^2$] & $310 \pm 6$ & $318 \pm 6$ & $233 \pm 5$ \\
$B$ [1/arcsec$^2$] & $29 \pm 1$ & $27 \pm 1$ & $11.7 \pm 0.6$ \\
$M$ [1/arcsec$^2$] & $11.89 \pm 0.09$ & $5.54 \pm 0.06$ & $0.77 \pm 0.03$ \\
\hline
\end{tabular}
\end{table}

With the fitted parameters of Eq.~\ref{equ: model distance} the probability $p$ of a match at the distance $r$ to be the true counterpart can be estimated through the expression
\begin{equation}
p(r) = \frac{A\cdot r \cdot \exp \left(-\frac{r^2}{2\sigma_{1}^2}\right) + B\cdot r \cdot \exp \left(-\frac{r^2}{2\sigma_{2}^2}\right)}{A\cdot r \cdot \exp \left(-\frac{r^2}{2\sigma_{1}^2}\right) + B\cdot r \cdot \exp \left(-\frac{r^2}{2\sigma_{2}^2}\right) + M \cdot r}.
\label{equ: probability function}
\end{equation}
Although the standard deviations $\sigma_1$ and $\sigma_2$ take the same value for all matching catalogs, the probability varies for a specific matching distance, because for a catalog with a lower source density one is less likely to find a random association. The differences in the source densities are caused by the magnitude ranges covered by the various catalogs. However, the matching probability does not consider the different brightnesses within a catalog. Therefore, the fraction of spurious identifications 
ought to be larger near the magnitude cutoff.

\label{sec: completeness and reliability}
\subsubsection{Missed and spurious identifications}
\label{sec: missed and spurious identifications}

The number of stellar identifications of the XMMSL2 sources missed in our sample, i.e., the completeness, and the number of spurious identifications in the sample, i.e., the reliability, depend on the chosen matching criteria. Especially the probability cutoff above which we assume matches to be the true counterpart controls these two characteristics (cf. Sect.~\ref{sec: matching probability}). The number of spurious identifications $N_\mathrm{spurious}$ in the sample with a constant matching distance $r$ can be estimated through
\begin{equation}
N_\mathrm{spurious}=\int\limits_0^r Mr' dr'.
\end{equation}
%where $M$ is the slope in the number distribution of the random associations (cf. Eq.~%\ref{equ: model distance}). 
In our sample the matching distance is not constant, but depends on the catalog used for the identification. The differential matching probability, defined in Eq.~\ref{equ: probability function}, gives the probability for a match at the distance $r$ to be the true counterpart. Hence we calculate the probabilities $p_i$ of each match up to a distance of $40$~arcsec where the probability to be the true counterpart drops to below 1\,\% for all our catalogs, i.e., these matches are negligible.

Given the probabilities $p_i$ and the chosen probability cutoff, the number of spuriously identified XMMSL2 sources can be estimated by summing the inverse probabilities of all sources with a probability higher than the cutoff $N_\mathrm{>cutoff}$. For a XMMSL2 source with $N_\mathrm{matches}$ matches, the probability that none of the matches is the true counterpart is given by the product of the inverse probabilities of all matches. Hence, we estimate the number of spuriously identified sources through 
\begin{equation}
N_\mathrm{spurious}=\sum\limits_i^{N_\mathrm{>cutoff}}\prod_j^{N_\mathrm{matches,i}}\left(1-p_j\right)
\label{equ: spurious identifications}
\end{equation} 
and we define the reliability $r$ of the matched sample as
\begin{equation}
r = \frac{N_\mathrm{>cutoff}-N_\mathrm{spurious}}{N_\mathrm{>cutoff}}.
\end{equation}
We estimate the number of missed identifications by summing the probabilities of all matches with a probability lower than the cutoff $N_\mathrm{<cutoff}$
\begin{equation}
N_\mathrm{missed} = \sum\limits_i^{N_\mathrm{<cutoff}} p_i.
\label{equ: missed identifications}
\end{equation} 
and define the completeness $C$ as 
\begin{equation}
C = \frac{N_\mathrm{>cutoff}}{N_\mathrm{>cutoff}+N_\mathrm{missed}}.
\end{equation}

\subsection{Associations in multiple catalogs}
\label{sec: Associations in multiple catalogs}

Quite a few XMMSL2 sources have counterparts in several catalogs, therefore we must determine whether we are considering the same counterpart or not. In some catalogs the identifications of other catalogs are specified; for example, the TGAS sample of the Gaia DR1 contains the Tycho2 identifier, which is, in this specific case, based exclusively on apparent sky distance. If no identifier is specified, we associate counterparts if their distance is smaller than $1$~arcsec or their extrapolated $V$ band magnitude difference is smaller than $1.5$~mag and their distance is smaller than $4$~arcsec. The exact distances do not influence the result significantly. We choose the closest match if these conditions are met by multiple counterparts. For measurements given in multiple catalogs, we use the value of the catalog with the highest accuracy of the respective measurement. 
To estimate the probability of a random match, we use the catalog with the highest probability.

\subsection{Additional stellar properties}
\label{sec: Estimation of additional stellar properties}

With the magnitudes in the available photometric bands, we estimate the effective temperatures, the bolometric magnitudes and fluxes as well as the $V$ band magnitudes for the sources with a 2MASS and Gaia counterpart only. We adopt the relations in Table~3 of \citet{color-table}, applying solar metallicity and a surface gravity of $g=10^{4.5}~\mathrm{cm\,s^{-2}}$ corresponding to dwarfs and use a linear interpolation of the values given in the table. We do not perform any corrections for extinction. Since \citet{color-table} provide the colors in the photometric system of Bessel \& Brett, we apply the relation given by \citet{2MASS-Photo} and in ''Explanatory Supplement to the 2MASS All Sky Data Release and Extended Mission Products'' \footnote{\url{http://www.ipac.caltech.edu/2mass/releases/allsky/doc/explsup.html}} to obtain the 2MASS colors. We further adopt the correlation to the Gaia band from \citet{Gaia-Photo} and the ''Gaia Data Release Documentation''\footnote{\url{https://gaia.esac.esa.int/documentation/GDR1/}}. 

For the calculation we adopt the $V-K$ color for the sources with a counterpart in the Tycho2 catalog and the 2MASS catalog, while we use the $B-V$ color and the $J-K$ color for the sources with a counterpart only in the Tycho2 catalog or the 2MASS catalog, respectively. We cannot estimate the bolometric flux and the effective temperature for a few sources ($\sim 2\%$), because they have a Gaia counterpart only or they are extremely red and lie outside the region defined in Table~3 of \citet{color-table}.

We find a trigonometric parallax for 57~\% of the stellar XMMSL2 sources and for these sources, we further estimate the distance, the $V$ band and bolometric absolute magnitude as well as the bolometric and the X-ray luminosity.

\subsection{Cleaning procedures}
\label{sec: cleaning procedures}

In our catalog we introduce different XMMSL2 source flags, if the sources have measurements of low quality or to indicate likely non-stellar objects. A few object classes will be identified as stellar sources by the procedures outlined above, 
that are not the focus of this work; examples are High- and Low Mass X-ray binaries, where
the X-ray emission is not predominantly produced by the star, but by matter accreted onto a compact object. Many of these objects
are already excluded because they generally have faint optical counterparts due to their very large X-ray/optical flux ratios, but
we additionally exclude sources which have a known accreting object in the SIMBAD database \citep{SIMBAD-database} within a distance of $30$~arcsec to the XMMSL2 detection. 
Furthermore, we flag sources that have a known galaxy cluster within $60$~arcsec or an active galactic nucleus (AGN) located within $30$~arcsec as listed the SIMBAD database. For such X-ray sources, the stellar object and the extragalactic object are both plausible counterparts, given the available information.

We additionally flag all sources that are not detected in all 2MASS bands and that have no association in other catalogs. Furthermore, the stellar identifications are unreliable, if they are flagged as extended in the 2MASS catalog or if they are detected in the XMMSL2 hard band only.
We expect sources to be affected by optical loading (i.e., X-ray events created by an excess of optical photons in the pn camera) if they are flagged in the XMMSL2 catalog and do not have a RASS counterpart (cf. Sect.~\ref{sec: RASS counterparts}). Some sources have erroneous 2MASS photometry or an unusual color, and very likely the derived magnitudes and stellar properties are unreliable.
Additionally, we flag sources with a high X-ray to the bolometric flux ratio. Specifically, we flag sources with a high Fx/Fbol ratio only in the slew survey and set an additional flag for sources that consistently have a high Fx/Fbol ratio in both, the slew and the RASS surveys and another flag is used for sources without RASS counterpart.

All used flags are summarized in Appendix~\ref{sec: flowchart} and \ref{sec: column description}. We generally exclude flagged sources from our subsequent analysis, but discuss some of their properties in the following sections.

\section{Results} 
\label{sec: results}

\subsection{Stellar counterparts, completeness, and reliability}
\begin{figure}[t]
%\captionsetup{labelfont=bf}
\resizebox{\hsize}{!}{\includegraphics{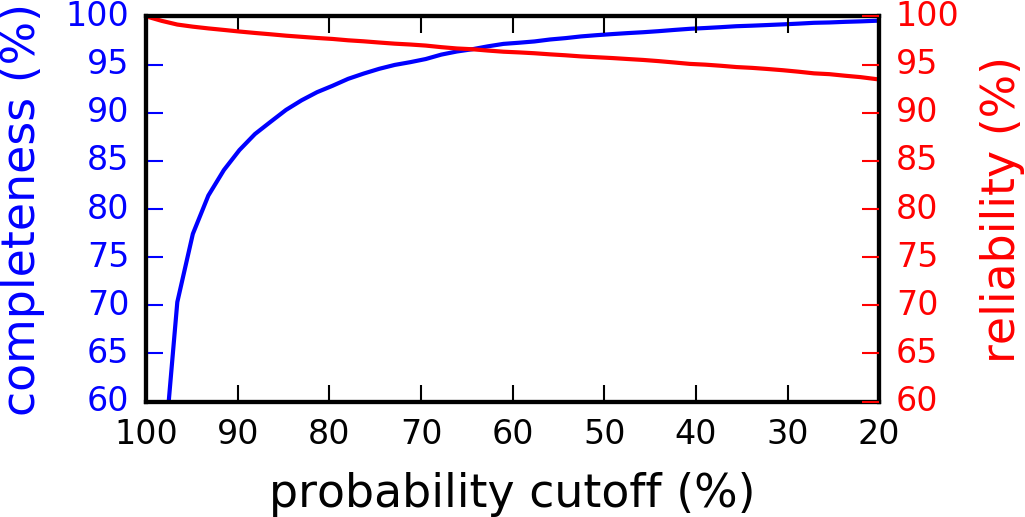}}
\caption{Completeness (blue curve) and reliability (red curve) of the sample as a function of the probability cutoff.}
\label{fig: completeness distribution}
\end{figure}

Figure~\ref{fig: completeness distribution} shows the completeness and reliability 
of the sample and its dependence on the matching probability for the non-flagged sources.
To balance completeness and reliability, we choose a matching probability $>2/3$, i.e., about intersection of the two curves, to derive the stellar catalog of XMMSL2 sources and obtain
a completeness of 96.3\,\% (Eq.~\ref{equ: missed identifications}) and a reliability of 96.7\,\% (Eq.~\ref{equ: spurious identifications}) ignoring sources that are flagged by our cleaning procedure in the calculations.
In terms of matching radii, this corresponds to $10.8$~arcsec, $13.9$~arcsec and $19.9$~arcsec for sources with a counterpart in the Gaia, 2MASS and Tycho2 catalogs, respectively.

Combing these matching radii with our adopted magnitude cutoffs, we find at least one stellar counterpart for 6815 of the 23\,252 XMMSL2 sources. 
Our cleaning procedures reduce the sample to 5920 sources, implying that 25.5 \% of the XMMSL2 sources are stellar sources.

\subsection{Single and multiple counterparts}
\label{sec: multiple counterparts}

\begin{figure}[t]
%\captionsetup{labelfont=bf}
\resizebox{\hsize}{!}{\includegraphics{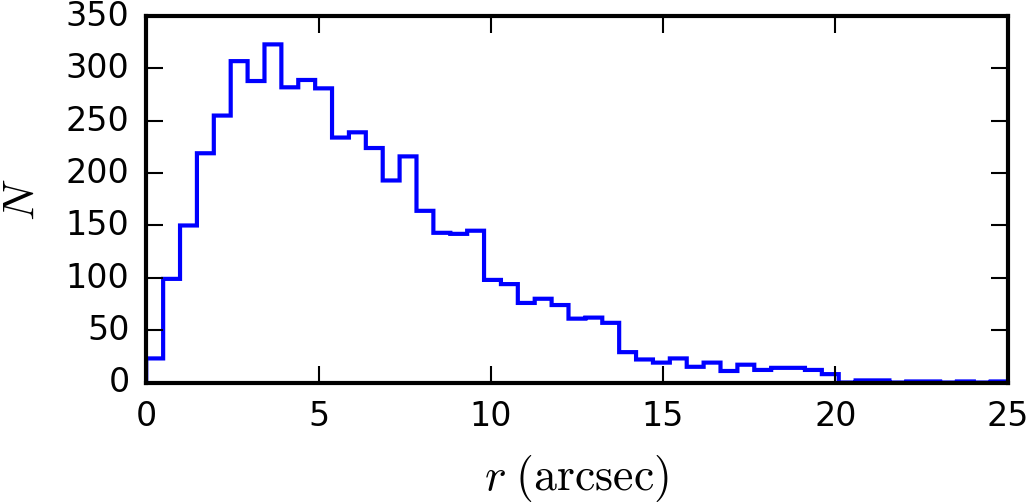}}
\caption{Distances between the XMMSL2 sources and the counterparts for the 5042 XMMSL2 sources with a single match.}
\label{fig: Distance single match}
\end{figure}

We specify the number of selected stellar counterparts per XMMSL2 source in Table~\ref{tab: Number of mutiple counterparts}.  
For the 5042 XMMSL2 sources with a single counterpart we present the histogram of the distances between the XMMSL2 sources and the counterparts in Fig.~\ref{fig: Distance single match}.
No strong discontinuities are visible at the adopted matching distances, only at $\approx 14$~arcsec, i.e for 2MASS identification without a Tycho2 counterpart, a slight drop is visible.
Sources with a distance $>14$~arcsec have a Tycho2 identification, and the few sources at a distance larger than $20$~arcsec have a counterpart in the BrightStar or Lepine catalog that we include, because of the small source density of these catalogs.

\begin{table}[t]
\centering
\caption{Number of multiple counterparts}
\label{tab: Number of mutiple counterparts}
\begin{tabular}{ll}
\hline \hline
Number of stellar counterparts & Number of sources \\
\hline
1 & 5042 \\
2 & 761 \\
3 & 103 \\
$>3$ & 14 \\
\hline
\end{tabular}
\end{table}

\begin{figure}[t]
%\captionsetup{labelfont=bf}
\resizebox{\hsize}{!}{\includegraphics{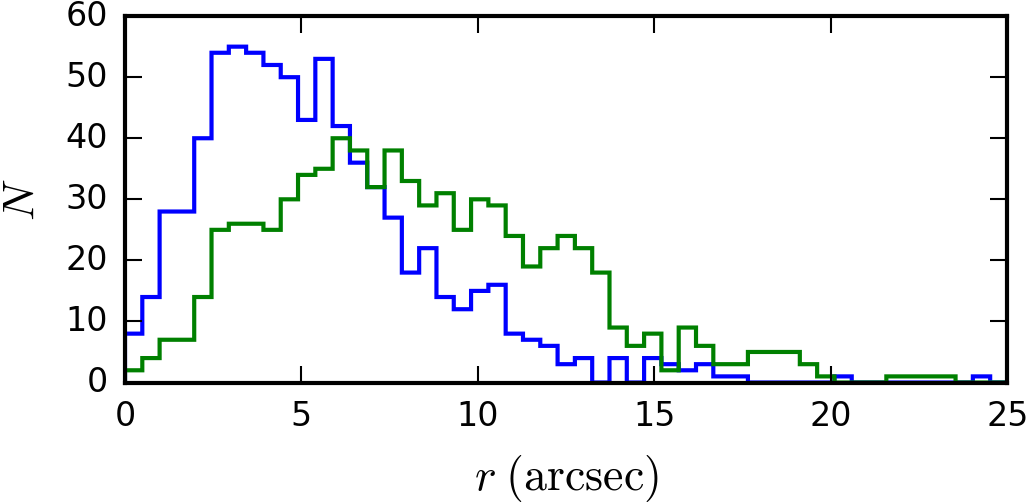}}
\caption{Distances between the XMMSL2 sources and the counterparts for the 761 XMMSL2 sources with two matches. The blue curve shows the distance of the sources with the higher matching probability; the green curve of the sources with the lower matching probability.}
\label{fig: Distance double match}
\end{figure}

\begin{figure}[t]
%\captionsetup{labelfont=bf}
\resizebox{\hsize}{!}{\includegraphics{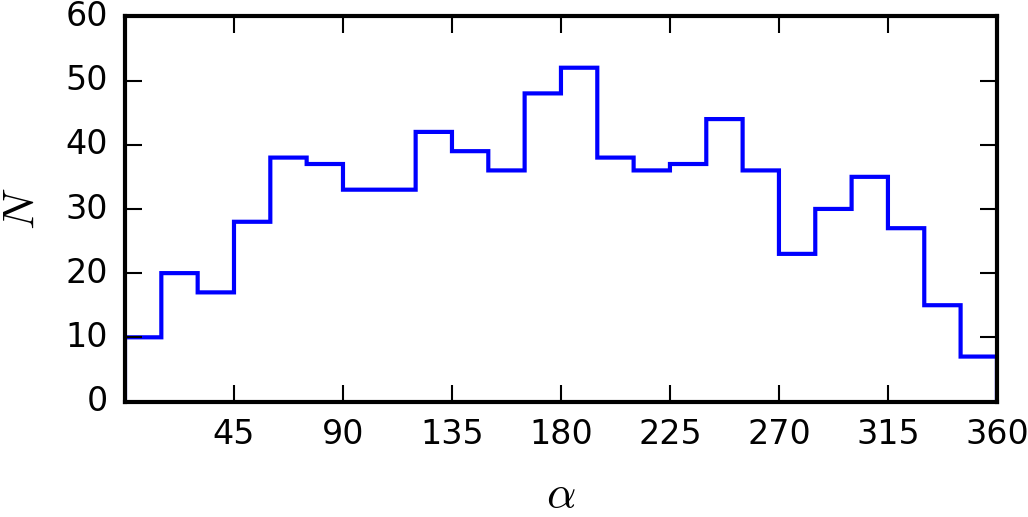}}
\caption{Distribution of the angle between the XMMSL2 source and the counterpart for the 761 sources with two counterparts.}
\label{fig: angle between counterparts}
\end{figure}

We find that roughly 15\,\% of the stellar XMMSL2 sources have more than one plausible stellar counterpart.
In Fig.~\ref{fig: Distance double match} we show the distances between the XMMSL2 sources and the counterparts for the 761 XMMSL2 sources with two matches. 
The mean angular separation of the counterparts with the higher matching probability for each XMMSL2 sources is smaller, but overall the two distributions are quite similar, i.e., both distributions have a maximum at small distances and the number of sources decreases for larger distances.
In Fig.~\ref{fig: angle between counterparts} we show the histogram of the angle between the XMMSL2 source and the counterparts for the sources with two counterparts. The distribution has a maximum at $180^\circ$, which implies that the XMMSL2 source tends to lie between both candidates. Thus, the cataloged XMMSL2 source is likely often a combination of the X-ray emissions of the two sources. We ignore sources with multiple counterparts when investigating the properties of the stellar counterparts.

\subsection{Catalogs of the counterparts}
\label{sec: catalogs of the counterparts}
In Table~\ref{tab: Catalogs of the counterparts} we denote the catalogs providing of the most probable counterpart.
Most counterparts have a Gaia and 2MASS identification as expected with 490 counterparts missing in the Gaia DR1 catalog. This is likely caused by the known incompleteness of the Gaia DR1 catalog. Furthermore, some of the Gaia sources cannot be identified with their 2MASS counterparts, possibly due observing epochs
that differ by about 15 years and unknown proper motion.
107 counterparts do not have a 2MASS counterpart. 
Multiple stars that are resolved in the Tycho2 or Gaia catalog, but not in 2MASS, could explain most of the 107 sources that do not have a 2MASS counterpart.
The two sources denoted as 'other' in Table~\ref{tab: Catalogs of the counterparts} only have a counterpart in the Lepine or BrightStar catalog with extremely high proper motions ($>4~\mathrm{arcsec\,yr^{-1}}$) and therefore, the 2MASS and Gaia counterparts, having no proper motion, lie outside of our initial matching radius of $40$~arcsec (cf. Appendix~\ref{sec: flowchart}).

\begin{table}[t]
\centering%
\caption{Catalogs of the counterparts}
\label{tab: Catalogs of the counterparts}
\begin{tabular}{lc|lc}
\hline\hline
Catalog & N & Catalog & N \\
\hline
TMG & 3595 & M & 116\\
MG & 1784 & T & 54 \\
TM & 318 & G & 15\\
TG & 36 & other & 2\\
\hline
\end{tabular}
\tablefoot{G: Gaia, M: 2MASS, T: Tycho2}
\end{table}

\subsection{Comparison to the identifications of \citet{sax08}}
\label{sec: identifications of Saxton}

The XMMSL2 catalog provides identifications and classifications for about 70\,\% of the X-ray sources as a result of a crossmatch with the SIMBAD, NED and
other databases and catalogs. However for some sources the classification only contains the region of the electromagnetic spectrum where the source has been detected, e.g., ''X-ray'', and provides little or no insight into the physical nature of the source. The classification adopted in the XMMSL2 catalog \citet{sax08} uses different catalog resources and enables an independent comparison to our results.

There is large overlap in the identification with 4231 sources consistently classified as stellar (the XMMSL2 catalog identifies a total of 5094 sources as stars). Our stellar identification is not confirmed for 1689  sources, however the vast majority (1671) of these has either no identification or the classification contains only the region of the electromagnetic spectrum. 
The XMMSL2 contradicts our stellar identification only for 18 (0.4\,\%) sources, typically referring to an additional plausible counterpart.

\begin{figure}[t]
%\captionsetup{labelfont=bf}
\resizebox{\hsize}{!}{\includegraphics{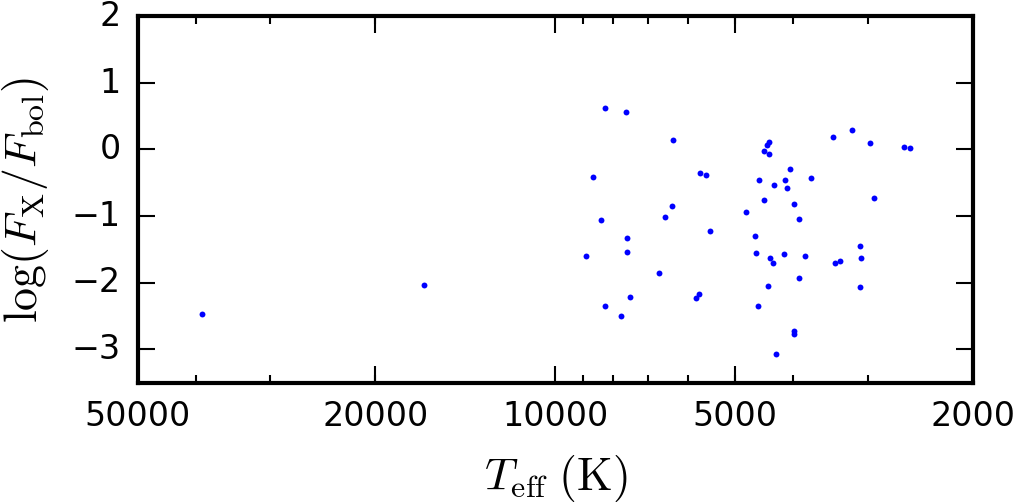}}
\caption{$F_\mathrm{X}/F_\mathrm{bol}$ distribution as a function of effective temperature of the sources classified as stellar in the XMMSL2 catalog, but missing a stellar counterpart with our approach.}
\label{fig: F_x-F_bol only XMMSL2 stars}
\end{figure}

On the other hand, 863 sources are classified as stars in the XMMSL2 catalog, where we do not find a stellar counterpart satisfying our selection criteria. 
Out of these, 531 are excluded by our cleaning procedures, which argues against a true stellar identification.
To validate the reliability of the remaining 332 stellar identifications, we inspect the $\log F_\mathrm{X}/F_\mathrm{bol}$ ratio as a function of the effective temperature for the 59 sources where colors from the SIMBAD database are available (see Fig.~\ref{fig: F_x-F_bol only XMMSL2 stars}).
For most of these sources we find $\log F_{\rm X}/F_{\rm bol} > -2$ or $T_{\rm eff} > 10\,000$~K and $\log F_{\rm X}/F_{\rm bol} > -3$, i.e., highly unlikely values for stellar X-ray sources. Overall, only a few of these sources remain as plausible stellar counterparts that are missed due to our chosen magnitude cutoff.   
Based on this fraction, we estimate that about 1\,\% of the stellar counterparts are missed by our procedure due to the applied brightness limit.

\subsection{Validation of our procedure}

\subsubsection{Application of our procedure to the EMSS catalog}
\label{sec: Application of our procedure to the EMSS catalog}
We validate the reliability of our identification approach by applying our procedures to the Extended Medium-Sensitive Survey  \citep[EMSS, ][]{gio90,sto91}. 
The EMSS catalog contains 835 X-ray sources detected in 1435 pointings with the Imaging Proportional Counter (IPC) on board the {\it Einstein Observatory} at high galactic latitudes ($|b|>20^\circ$). More than 96\,\% of the EMSS sources have been individually identified with spectroscopically classified counterparts. The reliability of these identifications should be very high, therefore the EMSS catalog is an ideal tool to test the reliability of our automatic procedure.

To identify the stellar content of the EMSS, we use basically the same procedure as for the XMMSL2 catalog, but we find a single Gaussian plus linear curve to adequately describe the matching distance distribution of the EMSS sources. Furthermore we search in the SIMBAD database for accreting and extragalactic objects within $50$~arcsec, the typical 90\,\% confidence error circle radius of the EMSS sources. 
%Out of these 6 are flagged as possible extragalactic objects, in the total sample 23 are flagged. 

%\subsubsection{Comparison to the classifications of the EMSS catalog}

We identify 210 of the 835 EMSS sources as stellar, whereas we expect about 15 identifications to be spurious and a similar number to be missed by the probability cutoff. These numbers can be compared with the identifications of the EMSS catalog, which contain 217 stars. 192 sources are consistently classified by the EMSS catalog and by our procedure.

The EMSS catalog thus identifies 25 sources as stars that we do not find with our procedure as stellar sources. One of these sources is marked as a white dwarf and 7 as cataclysmic variables.
Hence, 17 sources are stellar X-ray sources that are missed by our procedure, 10 because of the large distance to the EMSS source, 6 because they have a magnitude of $J>12$~mag and one case is uncertain. Note that we apply the same magnitude cutoff to the EMSS catalog as for the XMMSL2 catalog, which is a simplification and not an optimal cutoff for all EMSS sources. 
Hence, slightly fewer sources are missed because of the probability cutoff than expected.

For 18 sources the EMSS classification explicitly contradicts our stellar identification. However, for most of these sources there are two plausible counterparts in the SIMBAD database, i.e., one stellar counterpart and one extragalactic counterpart. So the stellar identifications maybe random associations, supported by the number of flagged sources in the total sample (23/210) copmared to consistently identified stars (6/192).
The number of 18 random associations is comparable to the expectation of 15 spurious identifications.

\begin{figure}[t]
%\captionsetup{labelfont=bf}
\resizebox{\hsize}{!}{\includegraphics{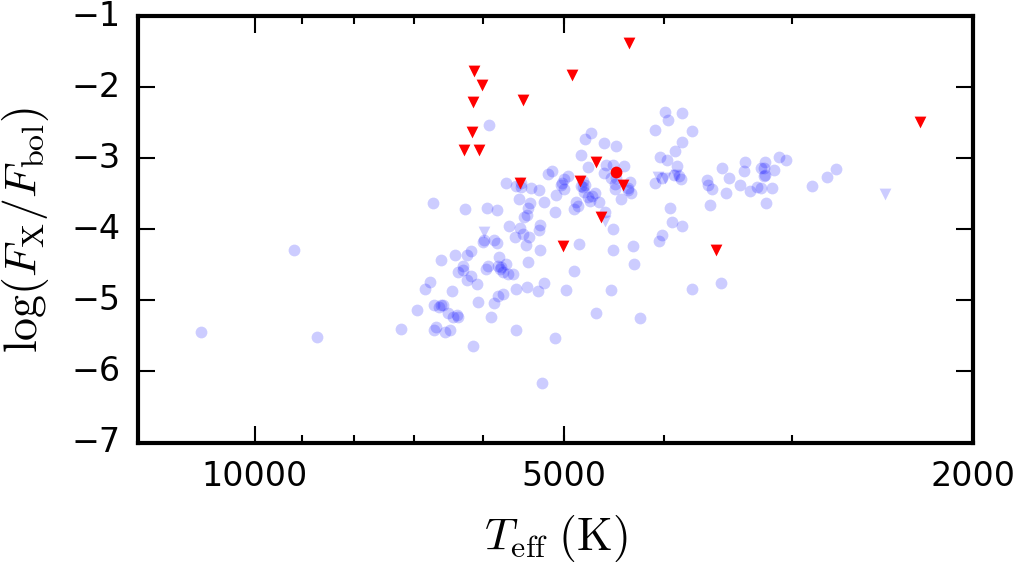}}
\caption{$F_\mathrm{X}/F_\mathrm{bol}$ distribution of the EMSS sources as a function of effective temperature. The blue symbols represent sources consistently identified as stars, red ones show sources classified as stars by our procedure, but with contradicting EMSS classification. The triangles denote possible extragalactic sources.}
\label{fig: F_x-F_bol EMSS}
\end{figure}

In Fig.~\ref{fig: F_x-F_bol EMSS} we present the $\log F_\mathrm{X}/F_\mathrm{bol}$ distribution of the 210 EMSS sources that we identify as stellar sources. Obviously, aboult half of the sources that have a contradicting classification in the EMSS catalog lie above the distribution of the sources that are consistently classified as stars. The other are at least plausible stellar X-ray sources, but we find that
they have generally a high angular separation and further they are mostly flagged as having a plausible extragalactic counterpart.
Therefore these stellar counterparts have a high chance of being random associations.

Yet, in summary the completeness and reliability of our automatic identifications reach the expectations. Therefore we are confident that the reliability and completeness calculated in Sect.~\ref{sec: missed and spurious identifications} represent the true reliability and completeness of the stellar XMMSL2 sample.

\subsubsection{Matching with the Chandra Source Catalog}
In comparison to XMM-Newton, the \textit{Chandra X-ray Observatory} \citep[CXO;][]{Chandra-mission1, Chandra-mission2} provides more accurate positions of the X-ray sources due to its sub-arcsec on-axis point spread function (PSF). However, in its current release 1.1 the Chandra Source Catalog \citep[CSC;][]{Chandra-catalog} covers less than 1~\% of the sky. Nevertheless, the precise positions of the CSC sources gives us the opportunity to validate some of our stellar identifications. Therefore, we perform a crossmatch of our stellar identifications with the CSC, applying a matching distance of 60~arcsec. Although the CSC sources generally have highly accurate positions, for some sources the positional uncertainty is much larger, and hence, we exclude sources with a 95~\% confidence error circle radius larger than 5~arcsec from our analysis. 

\begin{figure}[t]
%\captionsetup{labelfont=bf}
\resizebox{\hsize}{!}{\includegraphics{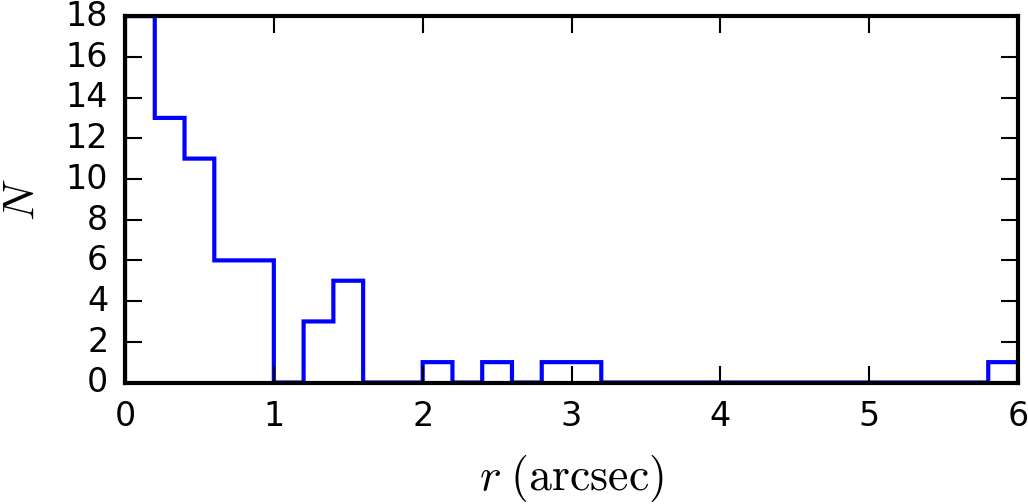}}
\caption{Angular separation between the stellar identification and the closest CSC counterpart for the XMMSL2 sources with a single match.}
\label{fig: Distance Chandra}
\end{figure}

In this fashion we find a CSC counterpart for 94 of the 5920 stellar identifications and show the angular separation between the stellar source and the CSC counterpart for the XMMSL2 sources with a single stellar identification in Fig.~\ref{fig: Distance Chandra}. For most of the sources (86) the distance between the CSC source and our best stellar identification is $<2$~arcsec as expected for the CSC sources or a larger angular separation can be explained by an unusual high positional uncertainty of the CSC source. For four sources the position of the CSC source indicates that our second best stellar counterpart is the correct identification of the X-ray source, and in four cases none of our stellar identifications lies within the 95~\% confidence error circle of the CSC source.  

%Furthermore, we find a CSC identification for 18 of the 761 sources with 2 stellar counterparts. For most sources our best stellar identification is confirmed by the CSC source. However, for two sources our second best stellar counterpart seems to be the correct identification of the X-ray source according to the position of the CSC source, and in one case both stellar sources cannot be identified with the CSC source. 9 of the 117 sources with more than 2 stellar counterpart can be associated with a CSC source. For 2 of these sources the position of the CSC source indicates that our second or fourth best stellar counterparts is the correct identification of the X-ray source, but for the other sources our best stellar identification is confirmed. 

In summary, the crossmatch with the CSC confirms at least one of our stellar sources for 95.7~\% of the XMMSL2 sources, which is in good agreement with the reliability calculated in Sect.~\ref{sec: missed and spurious identifications}.

\subsection{Catalog release}
We release the catalog of the stellar XMMSL2 sources at VizieR. It contains our stellar identifications with a matching probability $>2/3$. While we discuss in this paper only the properties of the unflagged sources, the released catalog also includes the stellar counterparts that are flagged by our cleaning procedure. XMMSL2 sources with multiple stellar counterparts have multiple entries in our catalog, one entry for each counterpart. We describe all new columns of our catalog in Appendix~\ref{sec: column description}.

\section{RASS counterparts}
\label{sec: RASS counterparts}

Before turning to the physical properties of our stellar counterparts, we compare and crossmatch the results from the XMMSL2 with the ''Second ROSAT all-sky survey (2RXS) source catalog'' \citep[][hereafter: RASS catalog]{RASS-catalog}.  Since the ROSAT all-sky survey flux
limit is deeper \mbox{($\sim 2\times 10^{-13}~\mathrm{erg\, cm^{-2}\, s^{-1}}$)} than that of the XMM-Newton slew survey ($\sim 5\times 10^{-13}~\mathrm{erg\, cm^{-2}\, s^{-1}}$), one would naively expect that all XMMSL2 sources
should have RASS counterparts. However, we find a RASS counterpart within $60$~arcsec for only 75.2\,\% of the sources; the matching fraction increases for multiple detected XMMSL2 sources to 91.6\,\%.

\begin{figure}[t]
%\captionsetup{labelfont=bf}
\resizebox{\hsize}{!}{\includegraphics{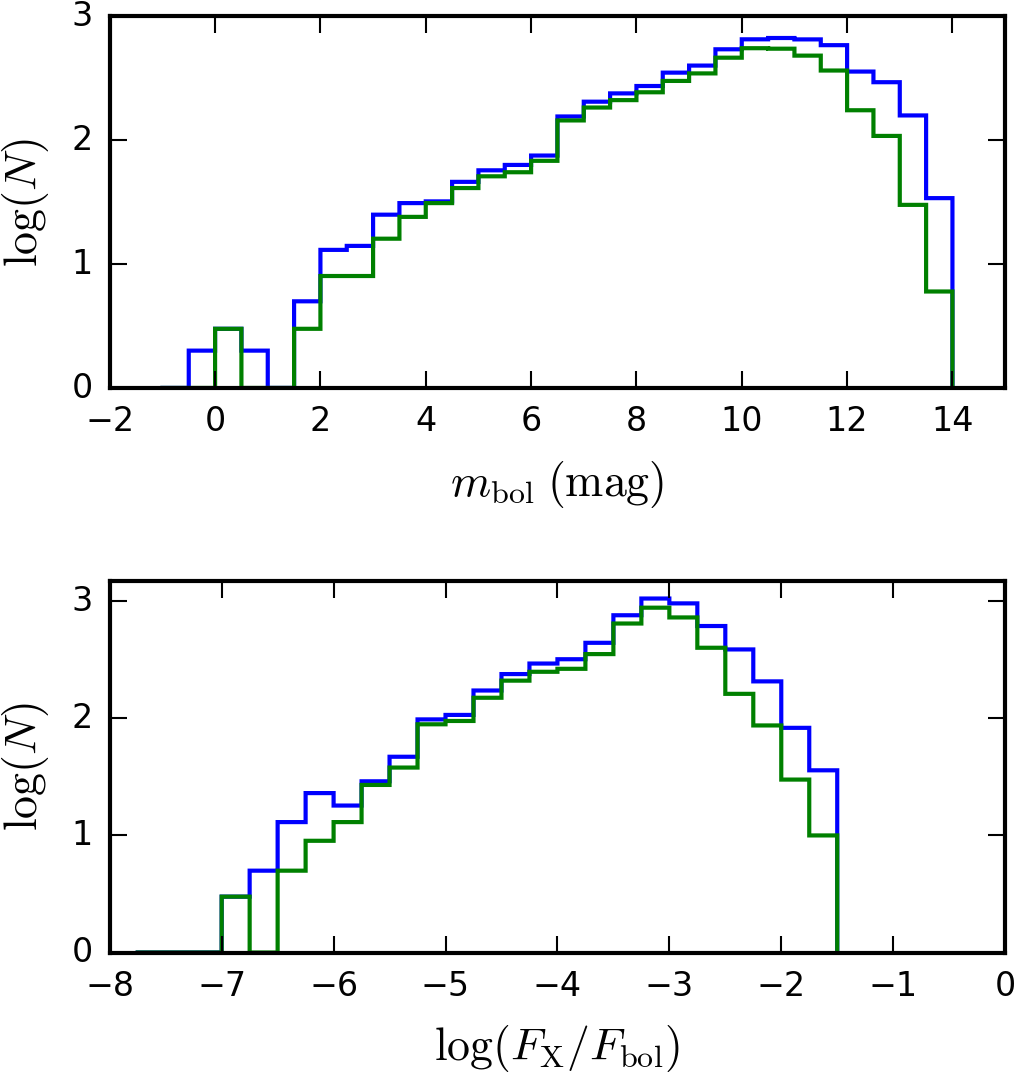}}
\caption{Number of all stellar XMMSL2 sources (blue) and of those with a RASS counterpart (green) as a function of {\it top}: apparent bolometric magnitude and {\it bottom}: X-ray activity.}
\label{fig: properties no RASS}
\end{figure}

To investigate the properties of the stellar XMMSL2 sources without a RASS identification, we compare their distribution as a function of the apparent bolometric magnitude and of the X-ray activity level to the full sample. As shown in Fig.~\ref{fig: properties no RASS}, the fraction of sources with a RASS identification is reduced for two different types of sources; first for bright sources, and second for highly active sources. For high activity stars we expect frequent flaring; thus, the small fraction of RASS identification is caused by the intrinsic variability of these stars.

At bright magnitudes we expect the small fraction of RASS identifications to be due to optical loading. 102 of the stellar XMMSL2 sources are flagged in the XMMSL2 catalog as possibly affected by optical loading. However, 68 of these sources do have a RASS identification and, hence, should be considered to be true X-ray emitters (RASS data is not affected by optical contamination). Furthermore, 6 sources brighter than $5$~mag without a RASS counterpart are not flagged. Therefore we conclude that the influence of optical loading cannot be reliably determined with the slew data only and hence, the optical loading flag should be used as an indicator together with additional information about the source. Note that the absolute number of XMMSL2 sources affected by optical loading is quite small and, therefore, intrinsic variability is the main reason why many stellar XMMSL2 sources do not have a 
RASS identification. We flag those stellar XMMSL2 sources that are marked as possibly affected by optical loading and that do not have a RASS identification (cf. Sect.~\ref{sec: cleaning procedures}).

\begin{figure}
%\captionsetup{labelfont=bf}
\resizebox{\hsize}{!}{\includegraphics{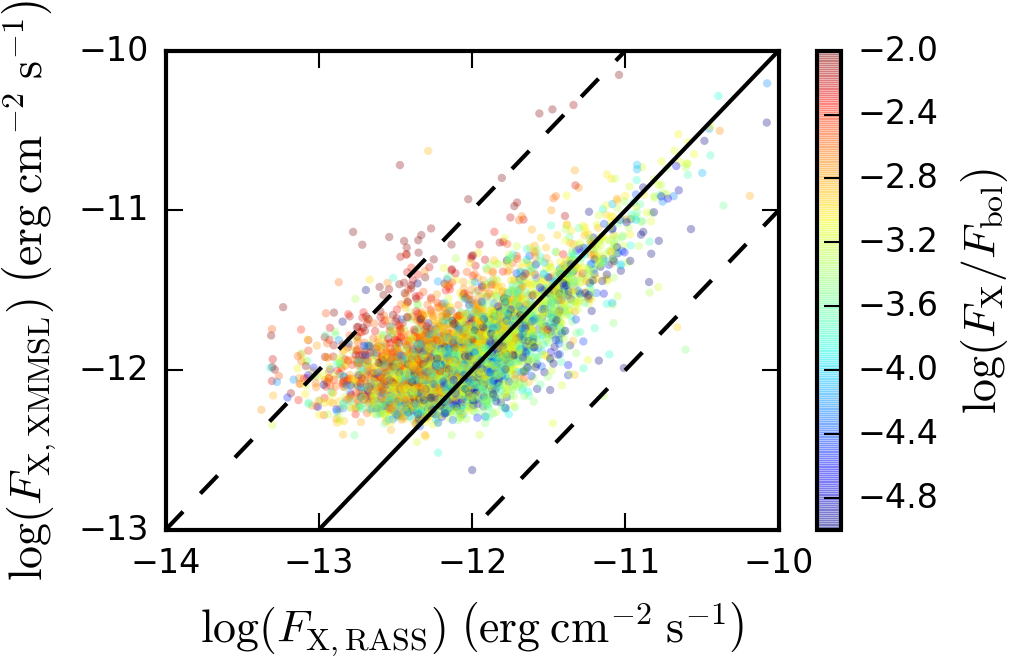}}
\caption{Comparison of the X-ray fluxes measured by XMM-Newton and ROSAT.  Solid line
indicates the same X-ray flux in the XMMSL2 and RASS catalogs, dashed lines
indicate flux difference of a factor of ten.
The color scales with the stellar X-ray activity measured by XMM-Newton.}
\label{fig: F_X RASS}
\end{figure}

For the XMMSL2 sources with a RASS counterpart two independent X-ray flux measurements 
are available. We convert the measured count rates of the XMMSL2 and RASS sources by applying the conversion factors defined in Sect.~\ref{sect: fluxes} and by \citet{sch95}, respectively. 
In Fig.~\ref{fig: F_X RASS} we compare the X-ray fluxes measured by XMM-Newton 
and ROSAT for the stellar XMMSL2 sources. For most sources the XMMSL2 flux is higher than the RASS flux, the median flux ratio is 1.4. The difference generally increases with increasing X-ray activity of the star as indicated by the color coding of Fig.~\ref{fig: F_X RASS}. 
For many XMMSL2 sources only 4-5 X-ray counts have been detected during the slew passages
and therefore, the uncertainty of the X-ray flux is quite high, although the
detection itself is very significant. This causes a rather strong bias for the flux level of sources at the detection limit, because many more sources lie just below the detection threshold than just above. Statistical fluctuations shift many of these above the threshold while the number of sources that have nominal fluxes above threshold, but remain undetected due to fluctuations to lower count numbers is considerably smaller. In effect, we inevitably overestimate the average X-ray flux of sources close to the detection threshold. Additionally, differences in the conversion factors might induce the systematically higher flux of the XMM then of ROSAT. Furthermore, we assume that the larger deviations are caused by intrinsic variability generated by flares. The RASS flux is less affected by flares, because the high flare fluxes have a smaller weight for RASS sources due to their longer exposure times.
Additionally some sources can only be detected during a flare in the XMMSL2 catalog, because the quiescent emission is below the XMMSL2 detection limit. In Fig.~\ref{fig: F_X RASS} the detection limit of the XMMSL2 catalog is visible at $\sim 5\times 10^{-13}~\mathrm{erg\, cm^{-2}\, s^{-1}}$, while a detection limit of the RASS catalog is not noticeable because in the RASS catalog the exposure time, and hence, the detection limit is not constant over the sky but the 
exposure time varies between typically $400$~s at the ecliptic equator and $\sim 40\,000$~s at the poles.

\section{Properties of the stellar sample}
\label{sec: properties}
Next we discuss the X-ray properties of the crossmatched stellar XMMSL2 sources. We also address the nature of the identified stellar counterparts, where we restrict the discussion to unique identifications, i.e. sources with exactly one stellar counterpart. However, in Sect.~\ref{sec: logN_logS} we use the stellar counterparts only to determine if the XMMSL2 source is stellar, and therefore, we also include sources with multiple counterparts. 

\subsection{Number-flux-density distribution}
\label{sec: logN_logS}

\begin{figure}
%\captionsetup{labelfont=bf}
  \resizebox{\hsize}{!}{\includegraphics{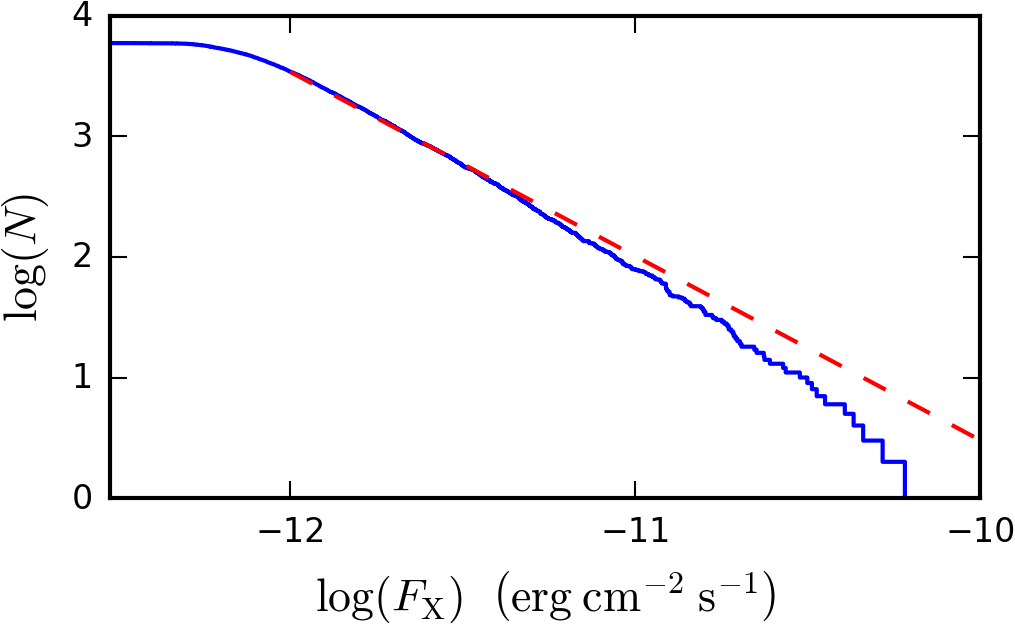}}
  \caption{Cumulative number-flux-density distribution of the stellar XMMSL2 sources. The red dashed line represents the best fit of the distribution.
  }
  \label{fig: LogN_logS}
\end{figure}

Figure~\ref{fig: LogN_logS} shows the derived number-flux-density distribution. 
At the X-ray bright end of the  diagram, the number of sources decreases approximately linearly with increasing flux in the double logarithmic scale and we describe the distribution by the power law ansatz
\begin{equation}
N(S) = k{F_\mathrm{X}}^{-\alpha},
\end{equation}
where $N$ is the number of sources with a flux brighter than $F_\mathrm{X}$, $\alpha$ is the slope, and $k$ is the normalization. Applying the method of \citep{cra70}, the best value of $\alpha$ can be estimated by maximizing the likelihood function
\begin{equation}
L = M\ln(\alpha) - (\alpha+1)\cdot\sum\limits_i \ln\left(\frac{F_{\mathrm{X,}i}}{F_\mathrm{X,min}}\right) - M\ln\left(1-\left(\frac{F_\mathrm{X,max}}{F_\mathrm{X,min}}\right)^{-\alpha}\right),
\end{equation}
where $M$ is the total number of sources, $F_\mathrm{X,max}$ and $F_\mathrm{X,min}$ are the brightest and faintest flux, respectively, and $F_{\mathrm{X,}i}$ is the flux of the i$^\mathrm{th}$ source. 

The slope depends on the lower flux limit imposed on the sample when fitting the distribution, because there is no hard detection limit due to different exposure times and background levels.
Using only sources brighter than $F_\mathrm{X}=10^{-12}$~erg$\,$cm$^{-2}\,$s$^{-1}$, we obtain a best value of $\alpha = 1.53 \pm 0.03$, which agrees to a spatially uniform distribution.

The best fit differs from the distribution of the brightest sources that show a slightly steeper slope. However, these sources have a smaller weight and higher uncertainty due to their small numbers. Most sources are found at faint fluxes (note the logarithmic scale in Fig.~\ref{fig: LogN_logS}), where a slope of $\alpha = 1.53$ fits the distribution well.

\subsection{X-ray luminosities}
\label{sec: X-ray luminosities}

\begin{figure}[t]
%\captionsetup{labelfont=bf}
\resizebox{\hsize}{!}{\includegraphics{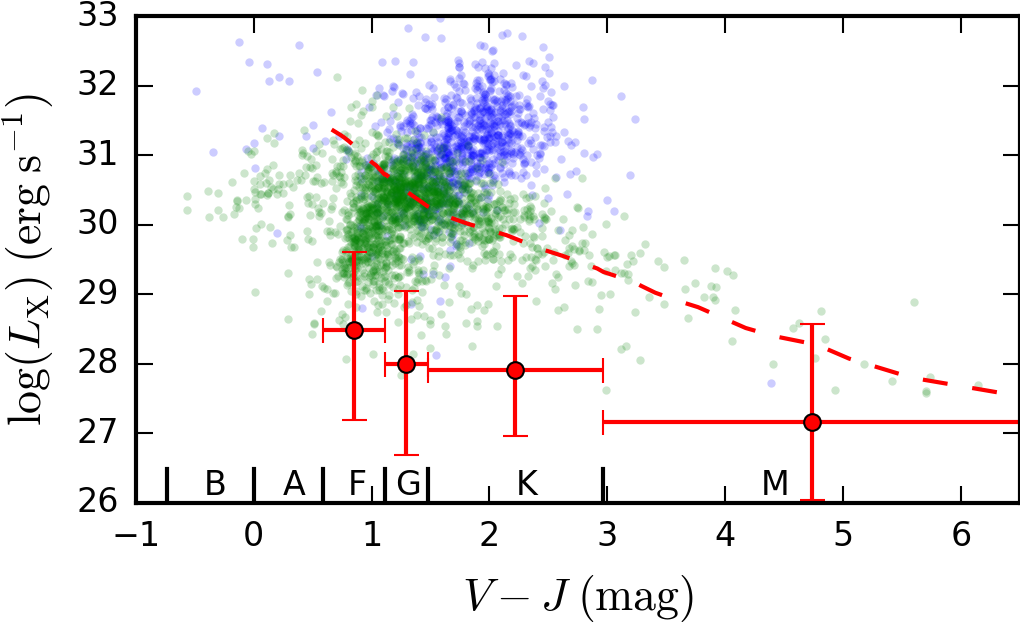}}
\caption{X-ray luminosity distribution of the stellar XMMSL2 sources with known parallax. Main sequence stars and giants are shown as green and blue dots, respectively. The red crosses show the X-ray luminosities of the volume limited NEXXUS sample \citep{schmitt04}. The extension in x-direction corresponds to the width of the spectral type, while the extension in y direction shows the luminosity range in which 90\,\% of the NEXXUS dwarfs of the specific spectral type lie. The red dashed line represents the saturation limit at $L_\mathrm{X}/L_\mathrm{bol}=10^{-3}$ for dwarfs.}
\label{fig: L_X distribution}
\end{figure}

In Fig.~\ref{fig: L_X distribution} we present the distribution of the X-ray luminosities as a function optical color
and compare the luminosities of our sample with those of the volume limited NEXXUS sample \citep{schmitt04}. The dwarf stars in our sample are up to two orders of magnitude brighter in X-rays than the most active NEXXUS sources of the same spectral type.
 Even the faintest sources in our sample are more luminous than the Sun, which is found at $\sim 3\times10^{26}~\mathrm{erg\,s^{-1}}$ and $\sim 5\times 10^{27}~\mathrm{erg\,s^{-1}}$ at solar minimum and maximum, respectively \citep{per00}. Hence, we only see the high luminosity tail of the stellar luminosity distribution in our sample. In addition, we expect many sources to be detected during a flare, which further biases our sample to high X-ray luminosities.

\subsection{Color-color diagram}
\label{sec: color-color diagram}
\begin{figure}[t]
%\captionsetup{labelfont=bf}
\resizebox{\hsize}{!}{\includegraphics{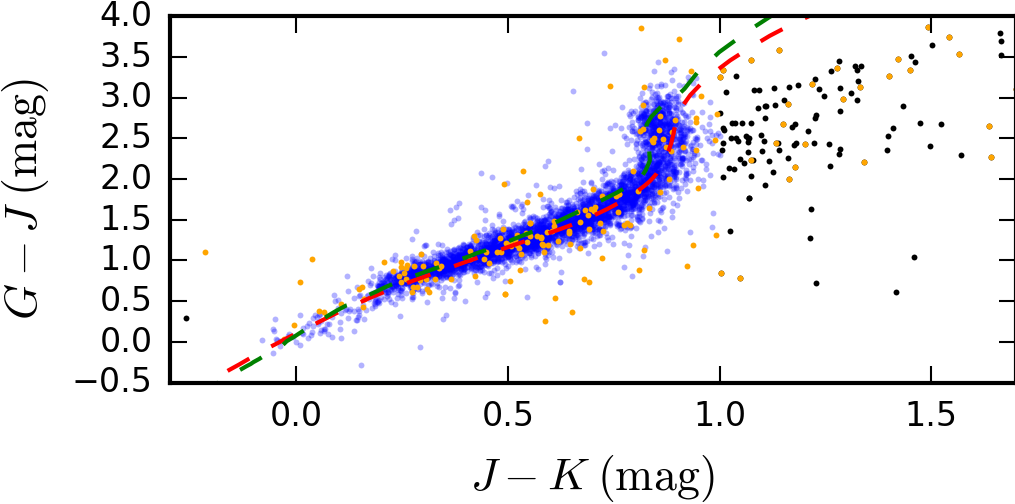}}
\caption{Color-color diagram of the stellar XMMSL2 sources with a Gaia and 2MASS counterparts. The sources represented as orange and black dots are flagged because of their known erroneous 2MASS photometry and their unusual colors, respectively. The red and green dashed curves represent different theoretical models of the main sequence; see text for details.}
\label{fig: color-color}
\end{figure}

About $80\,\%$ of the unique sources are covered by the 2MASS {\it and} the Gaia catalogs, and we construct color-color diagrams for those sources. Figure~\ref{fig: color-color} shows that the sources are arranged in a well defined streak. However, a few sources are located outside of the main distribution, many of them are known to have an erroneous 2MASS photometry. Many sources at the red end are classified as pre-main sequence stars by SIMBAD and their location in Fig.~\ref{fig: color-color} is likely due to reddening. 
Since we do not perform any correction for extinction (neither in the optical nor at X-ray wavelengths), we flag these sources in our catalog. Another possibility is an error in the optical magnitudes leading to incorrect colors, again, this motivates flagging these sources.

Figure~\ref{fig: color-color} also shows two theoretical color-color relations for main sequence stars. We adopt color-color relations from \citet{color-table} and \citet{pec13} to obtain the theoretical correlation between $G-J$ and $J-K$. 
The estimated theoretical color-color relation generally corresponds well with the observed distribution, and differs only for the sources with $J-K \approx 0.8$. However, for these sources the predicted correlation between $G-J$ and $J-K$ substantially depends on the assumed model.
 
\subsection{Hertzsprung-Russell diagram}
\label{sec: HRD}
\begin{figure*}[t]
%\captionsetup{labelfont=bf}
\sidecaption
\includegraphics[width=12cm]{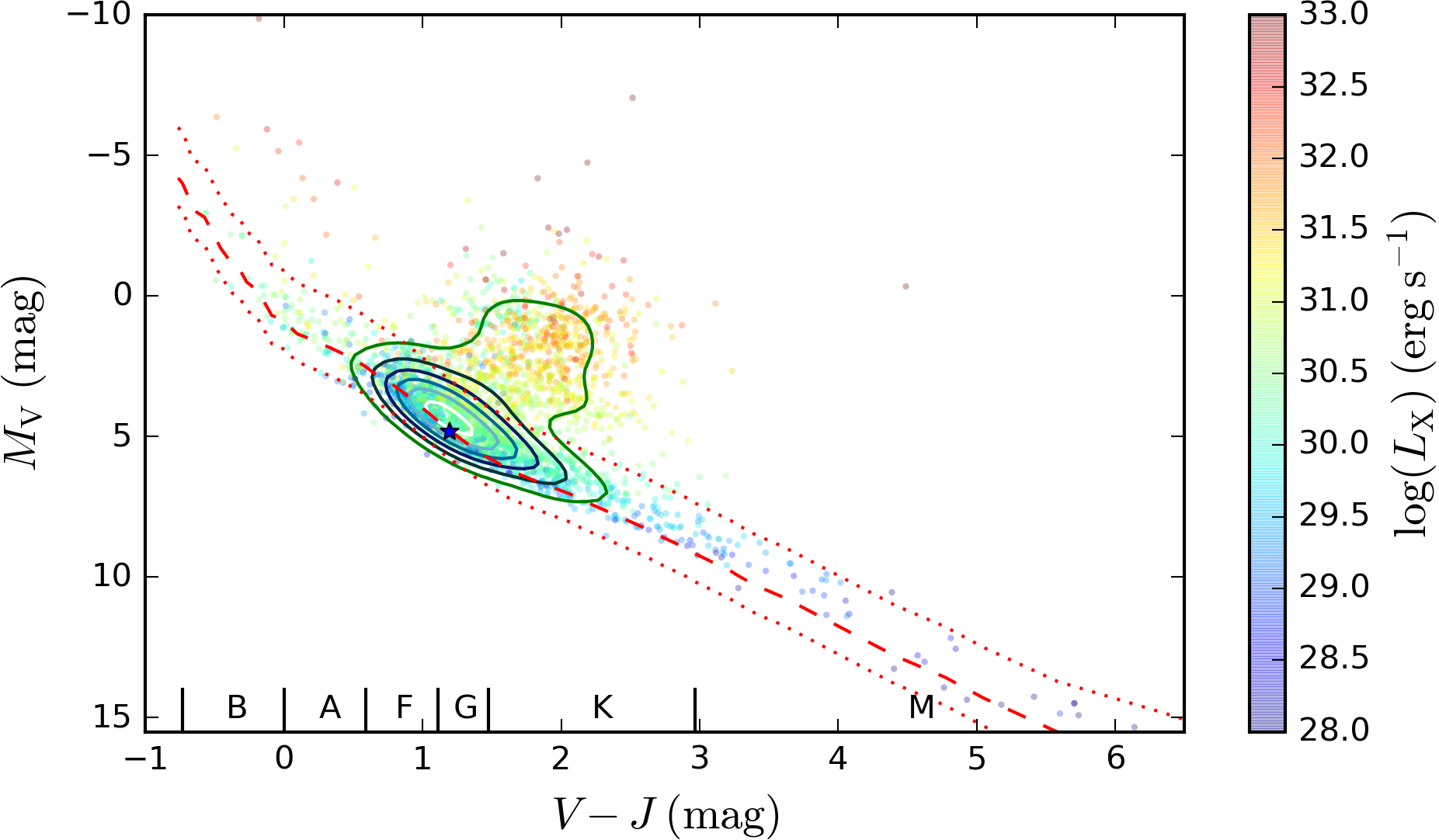}
\caption{Hertzsprung-Russell diagram of the stellar XMMSL2 sources with known parallax, the color scales with the logarithmic X-ray luminosity $\log(L_\mathrm{X})$. The red dashed line represents the theoretical main sequence following \citet{pec13}, and the dotted lines show the assumed width of the main sequence, the star marker represents the position of the Sun; note that the Sun is less luminous than the stellar XMMSL2 sources. The density of the distribution is indicated by the colored lines. The ranges of the spectral types for dwarfs are given at the bottom of the figure.}
\label{fig: HRD}
\end{figure*}

About 57\,\% of our sources with an unique stellar identification have trigonometric parallaxes, either from Gaia or the HIPPARCOS catalog \citep{hip97}, or in the Lepine and \mbox{BrightStar} catalogs.
Figure~\ref{fig: HRD} shows the Hertzsprung-Russell diagram (HRD) of these sources. 
We point out that
this sample of stellar counterparts with trigonometric parallaxes is neither complete in brightness nor in volume (many optically faint sources currently lack parallaxes),
however, Fig.~\ref{fig: HRD} does contain all basic features known from an optically selected HRD. The main sequence covers a wide range ($-0.9~\mathrm{mag}<V-J<6~\mathrm{mag}$), hence the sample contains all spectral types from O type stars down to dwarfs of spectral type M6V albeit sparsely populated at the limiting spectral types. 
Figure~\ref{fig: HRD} contains only a small number of M type dwarfs and some of them might actually be reddened K type dwarfs, although late-type dwarfs show frequent and extreme flares. However, the fluxes at X-ray and optical wavelengths are low for these sources, and hence, the parallaxes are currently unknown.  
Furthermore, the complete sample also includes even later type stars, but these sources lack parallaxes and are mainly flagged because of the extreme color.

In the following, we assume sources to be main sequence stars if their absolute brightness $M_\mathrm{V}$ is in the range $M_\mathrm{V,theo}+1.0>M_V>M_\mathrm{V,theo}-1.8$, where $M_\mathrm{V,theo}$ is the theoretical absolute brightness of a main sequence star adopted from \citet{pec13}. We find that 64\,\% of the sources are dwarfs. However, not all of these sources need to be single stars, rather we expect many sources to be X-ray bright, active binaries.
Also, the giant branch is clearly evident and we expect many of these sources 
to be RS~CVn systems. 

The HRD presented by \citet{gud04} contains about 2000 stars and
shares many similarities with our HRD. However, the number of stars associated with the different 
categories differs strongly, because our sample is drawn from a flux limited sample that 
is biased to active systems, while \citet{gud04} congregated data from several studies. This also explains why we only find a few early-type stars and no separate population of pre-main sequence stars.
Nevertheless, we expect that our complete sample contains some pre-main sequence stars, but these sources either lack parallaxes or are flagged because of reddening.

\subsection{$F_X/F_{bol}$-ratio}
\label{sec: F_x-F_bol}

\begin{figure*}[t]
\sidecaption
%\captionsetup{labelfont=bf}
\includegraphics[width=12cm]{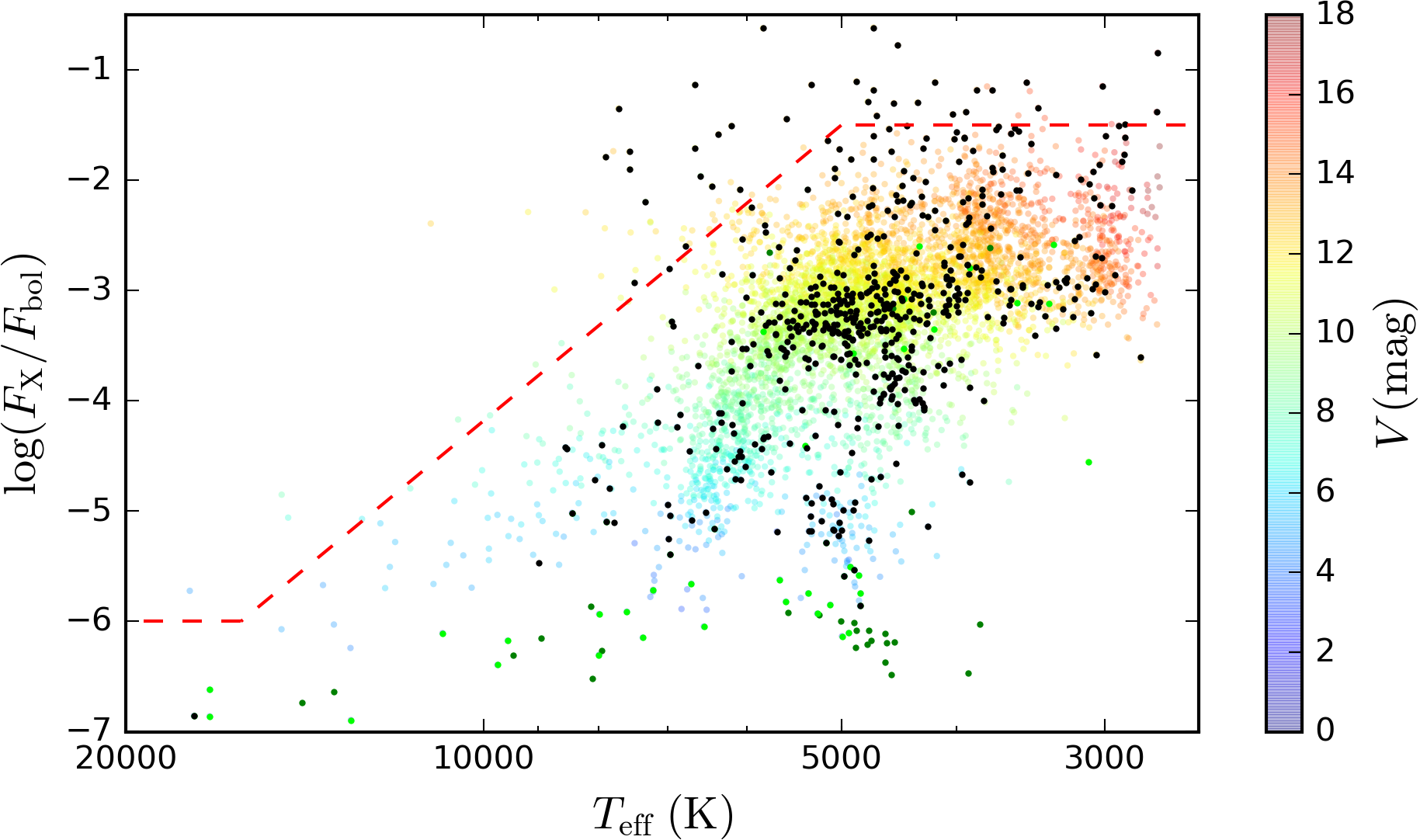}
\caption{$F_\mathrm{X}/F_\mathrm{bol}$ distribution of the stellar XMMSL2 sources as a function of  effective temperature. The colors scale with $V$ band magnitude. Sources that may be
affected by optical loading are shown as green dots, while those sources with a RASS identification are marked in light green color, and those sources without RASS counterpart are shown as dark green dots. The black dots represent highly reliable XMMSL2 sources with more than 17~cts (about 10\,\% of our sample). Sources lying above the red dashed line are flagged in our catalog as high X-ray activity during the slew.}
\label{fig: F_x-F_bol}
\end{figure*}

Figure~\ref{fig: F_x-F_bol} shows the $F_\mathrm{X}/F_\mathrm{bol}$ distribution of the stellar XMMSL2 sources as a function of the effective temperature (see Fig.~\ref{fig: properties no RASS} for the histogram of the distribution). We also include sources that are flagged in our catalog due to optical loading or their high $F_\mathrm{X}/F_\mathrm{bol}$ values.
The stars cover a wide range of activity with some sources having very high values of $F_\mathrm{X}/F_\mathrm{bol}$ and we find that about 14\,\% of the sources have an X-ray activity of 
$\log(F_\mathrm{X}/F_\mathrm{bol})>-2.5$, which is substantially higher than the saturation limit found in previous studies \citep[e.g.][]{pizzo03}. For many of these sources the RASS flux is up to 10 times fainter (see Fig.~\ref{fig: F_X RASS}) and we expect them to be detected by XMM-Newton during a flare.
We flag sources with persistent (XMMSL+RASS) high X-ray flux since these are probably non-stellar sources.
Furthermore, we flag strong transients and sources lying above the red dashed line in Fig.~\ref{fig: F_x-F_bol}; the high X-ray activity of these sources is unlikely 
caused by flares. Almost 6\,\% of the sources lie above the threshold of $\log(F_\mathrm{X}/F_\mathrm{bol})>-2.2$ that we use to calculate the magnitude cutoff. Since this magnitude cutoff is particularly relevant at both ends of the spectral type distribution, some true stellar identifications might be missed, especially in the M~dwarf regime. At very low masses, sources are flagged due to the extreme red colors and are missing in this figure.

The X-ray activity of the stars strongly depends on the spectral type, and
several known features are visible in Fig.~\ref{fig: F_x-F_bol}. First, the onset of convection at about $7000$~K is evident. Second, two distinct ''fingers'' at $7000$~K and $5000$~K contain sources down to relatively low activity levels. Most sources in the first finger are F dwarfs, while the second finger mainly consists of late G and early K giants. Since the X-ray flux has to be above the detection limit, the sources with a very low activity level must have a high (apparent) bolometric flux. It is also evident from the color coding that the low activity sources are relatively bright in the optical and are either nearby, inactive dwarfs or distant, X-ray bright giants. 
Third, there are few early type stars, which have a low X-ray flux compared to their bolometric flux. 

We show the sources with more than 17 counts (10\,\% of the samples) and, hence, with highly reliable X-ray fluxes as black dots in Fig.~\ref{fig: F_x-F_bol}. Most of these sources have $\log(F_\mathrm{X}/F_\mathrm{bol})<-3$, however, for some sources the contribution of the X-ray flux to the bolometric flux is extremely high. These sources could not be associated with a stellar counterpart brighter than the magnitude cutoff, if the source had been detected at the XMMSL2 detection limit.

\section{Conclusions and outlook}
\label{sec: Conclusions}
In this paper we present the first in-depth analysis of the stellar content of the 
XMM-Newton slew survey. In our approach the stellar XMMSL2 sources are identified by an automatic crossmatch 
of the XMMSL2 catalog with the Gaia DR1, 2MASS and Tycho2 catalogs;  we expect that in the future all necessary 
stellar data may actually be adopted from the new Gaia releases.  The
reliability of our automatic identification procedure is tested and verified by a comparison with the 
individually classified EMSS sources
and by a crossmatch of our stellar identifications with the CSC.

With our procedure a total of 6815 stellar counterparts are found for the 23\,252 XMMSL2 sources, after applying our cleaning procedures we identify 5920 XMMSL2 sources as high quality stellar X-ray sources corresponding to a stellar fraction of 25.5\,\%. 
We further expect that 195 stellar classifications are spurious, i.e., they are erroneously attributed to a star.  
On the other hand, we expect 227 stellar identifications to be missed due to the chosen probability cutoff. Therefore our sample has a reliability of 96.7\,\% and a completeness of 96.3\,\%. We further expect a small fraction of true stellar identifications to be missed due to our magnitude cutoffs. This is an significant 
improvement to the stellar classifications presented by \citet{sax08} (who had no Gaia data at their disposal), who identify only 4231 of our 5920 stellar XMMSL2 sources as stars. \citet{sax08} further give a stellar identification for 863 XMMSL2 sources that we do not identify as stars with our procedure, however, we show that most of the sources are less reliable stellar counterparts due to their high $F_\mathrm{X}/F_\mathrm{bol}$ values.

The XMMSL2 catalog contains stars of all spectral types and luminosity classes, however, most of the stellar XMMSL2 sources are -- not
unexpectedly -- late-type dwarfs with an outer convection zone. 
Only about 75\,\% of the XMMSL2 sources have a RASS identification, although the RASS catalog is deeper by a factor of about 2.5. Hence, a substantial portion of the stellar XMMSL2 sources are previously unknown X-ray sources caught in an active or flaring 
state.  With procedures as developed in this paper we expect to be able to 
perform an ''on-line'' identification of the stellar sources in the upcoming eROSITA
all-sky survey.

\begin{acknowledgements}
This research has made use of data obtained from XMMSL2, the Second XMM-Newton Slew Survey Catalogue, produced by members of the XMM SOC, the EPIC consortium, and using work carried out in the context of the EXTraS project ("Exploring the X-ray Transient and variable Sky", funded from the EU's Seventh Framework Programme under grant agreement no. 607452).
Additionally, we used data from the European Space Agency (ESA)
mission {\it Gaia} (\url{https://www.cosmos.esa.int/gaia}), processed by
the {\it Gaia} Data Processing and Analysis Consortium (DPAC,
\url{https://www.cosmos.esa.int/web/gaia/dpac/consortium}). Funding
for the DPAC has been provided by national institutions, in particular
the institutions participating in the {\it Gaia} Multilateral Agreement.
Furthermore, this publication made use of data products from the Two Micron All Sky Survey, which is a joint project of the University of Massachusetts and the Infrared Processing and Analysis Center/California Institute of Technology, funded by the National Aeronautics and Space Administration and the National Science Foundation.
Moreover, we used the VizieR catalogue access tool and the SIMBAD database, operated at CDS, Strasbourg, France. The original description of the VizieR service was published in A\&AS 143, 23 and we have made use of data obtained from the Chandra Source Catalog, provided by the Chandra X-ray Center (CXC) as part of the Chandra Data Archive. SF acknowledge supports through the Integrationsamt of Hamburg University, JR by DLR and CS through the SFB 676 founded by DFG.
\end{acknowledgements}

\bibliographystyle{aa} % style aa.bst
\bibliography{mybib}

\begin{thebibliography}{50}
\expandafter\ifx\csname natexlab\endcsname\relax\def\natexlab#1{#1}\fi

\bibitem[{{Agrawal} {et~al.}(1986){Agrawal}, {Rao}, \& {Sreekantan}}]{agr86}
{Agrawal}, P.~C., {Rao}, A.~R., \& {Sreekantan}, B.~V. 1986, \mnras, 219, 225

\bibitem[{{Ayres}(2009)}]{Ayres_2009}
{Ayres}, T.~R. 2009, \apj, 696, 1931

\bibitem[{{Berghoefer} {et~al.}(1997){Berghoefer}, {Schmitt}, {Danner}, \&
  {Cassinelli}}]{ber97}
{Berghoefer}, T.~W., {Schmitt}, J.~H.~M.~M., {Danner}, R., \& {Cassinelli},
  J.~P. 1997, \aap, 322, 167

\bibitem[{{Boller} {et~al.}(2016){Boller}, {Freyberg}, {Tr{\"u}mper}, {Haberl},
  {Voges}, \& {Nandra}}]{RASS-catalog}
{Boller}, T., {Freyberg}, M.~J., {Tr{\"u}mper}, J., {et~al.} 2016, \aap, 588,
  A103

\bibitem[{{Carpenter}(2001)}]{2MASS-Photo}
{Carpenter}, J.~M. 2001, \aj, 121, 2851

\bibitem[{{Catura} {et~al.}(1975){Catura}, {Acton}, \& {Johnson}}]{cat75}
{Catura}, R.~C., {Acton}, L.~W., \& {Johnson}, H.~M. 1975, \apjl, 196, L47

\bibitem[{{Crawford} {et~al.}(1970){Crawford}, {Jauncey}, \& {Murdoch}}]{cra70}
{Crawford}, D.~F., {Jauncey}, D.~L., \& {Murdoch}, H.~S. 1970, \apj, 162, 405

\bibitem[{{Dempsey} {et~al.}(1993){Dempsey}, {Linsky}, {Fleming}, \&
  {Schmitt}}]{dem93}
{Dempsey}, R.~C., {Linsky}, J.~L., {Fleming}, T.~A., \& {Schmitt}, J.~H.~M.~M.
  1993, \apjs, 86, 599

\bibitem[{{ESA}(1997)}]{hip97}
{ESA}, ed. 1997, ESA Special Publication, Vol. 1200, {The HIPPARCOS and TYCHO
  catalogues. Astrometric and photometric star catalogues derived from the ESA
  HIPPARCOS Space Astrometry Mission}

\bibitem[{{Evans} {et~al.}(2010){Evans}, {Primini}, {Glotfelty}, {Anderson},
  {Bonaventura}, {Chen}, {Davis}, {Doe}, {Evans}, {Fabbiano}, {Galle}, {Gibbs},
  {Grier}, {Hain}, {Hall}, {Harbo}, {(Helen He}, {Houck}, {Karovska},
  {Kashyap}, {Lauer}, {McCollough}, {McDowell}, {Miller}, {Mitschang},
  {Morgan}, {Mossman}, {Nichols}, {Nowak}, {Plummer}, {Refsdal}, {Rots},
  {Siemiginowska}, {Sundheim}, {Tibbetts}, {Van Stone}, {Winkelman}, \&
  {Zografou}}]{Chandra-catalog}
{Evans}, I.~N., {Primini}, F.~A., {Glotfelty}, K.~J., {et~al.} 2010, \apjs,
  189, 37

\bibitem[{{Favata} {et~al.}(2008){Favata}, {Micela}, {Orlando}, {Schmitt},
  {Sciortino}, \& {Hall}}]{Favata_2008}
{Favata}, F., {Micela}, G., {Orlando}, S., {et~al.} 2008, \aap, 490, 1121

\bibitem[{{Fleming} {et~al.}(1988){Fleming}, {Liebert}, {Gioia}, \&
  {Maccacaro}}]{fle88}
{Fleming}, T.~A., {Liebert}, J., {Gioia}, I.~M., \& {Maccacaro}, T. 1988, \apj,
  331, 958

\bibitem[{{Gaia Collaboration} {et~al.}(2016{\natexlab{a}}){Gaia
  Collaboration}, {Brown}, {Vallenari}, {Prusti}, {de Bruijne}, {Mignard},
  {Drimmel}, {Babusiaux}, {Bailer-Jones}, {Bastian}, \& et~al.}]{GaiaDR1}
{Gaia Collaboration}, {Brown}, A.~G.~A., {Vallenari}, A., {et~al.}
  2016{\natexlab{a}}, \aap, 595, A2

\bibitem[{{Gaia Collaboration} {et~al.}(2016{\natexlab{b}}){Gaia
  Collaboration}, {Prusti}, {de Bruijne}, {Brown}, {Vallenari}, {Babusiaux},
  {Bailer-Jones}, {Bastian}, {Biermann}, {Evans}, \& et~al.}]{Gaia-mission}
{Gaia Collaboration}, {Prusti}, T., {de Bruijne}, J.~H.~J., {et~al.}
  2016{\natexlab{b}}, \aap, 595, A1

\bibitem[{{Gioia} {et~al.}(1990){Gioia}, {Maccacaro}, {Schild}, {Wolter},
  {Stocke}, {Morris}, \& {Henry}}]{gio90}
{Gioia}, I.~M., {Maccacaro}, T., {Schild}, R.~E., {et~al.} 1990, \apjs, 72, 567

\bibitem[{{G{\"u}del}(2004)}]{gud04}
{G{\"u}del}, M. 2004, \aapr, 12, 71

\bibitem[{{Haisch} {et~al.}(1991){Haisch}, {Schmitt}, \& {Rosso}}]{haisch91}
{Haisch}, B., {Schmitt}, J.~H.~M.~M., \& {Rosso}, C. 1991, \apjl, 383, L15

\bibitem[{{Hempelmann} {et~al.}(2003){Hempelmann}, {Schmitt}, {Baliunas}, \&
  {Donahue}}]{Hempelmann_2003}
{Hempelmann}, A., {Schmitt}, J.~H.~M.~M., {Baliunas}, S.~L., \& {Donahue},
  R.~A. 2003, \aap, 406, L39

\bibitem[{{Hoffleit} \& {Jaschek}(1991)}]{BrightStar}
{Hoffleit}, D. \& {Jaschek}, C. 1991, {The Bright star catalogue}

\bibitem[{{H{\o}g} {et~al.}(2000){H{\o}g}, {Fabricius}, {Makarov}, {Urban},
  {Corbin}, {Wycoff}, {Bastian}, {Schwekendiek}, \& {Wicenec}}]{Tycho2}
{H{\o}g}, E., {Fabricius}, C., {Makarov}, V.~V., {et~al.} 2000, \aap, 355, L27

\bibitem[{{Huensch} {et~al.}(1996){Huensch}, {Schmitt}, {Schroeder}, \&
  {Reimers}}]{huensch96}
{Huensch}, M., {Schmitt}, J.~H.~M.~M., {Schroeder}, K.-P., \& {Reimers}, D.
  1996, \aap, 310, 801

\bibitem[{{Huensch} {et~al.}(1998{\natexlab{a}}){Huensch}, {Schmitt}, \&
  {Voges}}]{huensch98b}
{Huensch}, M., {Schmitt}, J.~H.~M.~M., \& {Voges}, W. 1998{\natexlab{a}},
  \aaps, 127, 251

\bibitem[{{Huensch} {et~al.}(1998{\natexlab{b}}){Huensch}, {Schmitt}, \&
  {Voges}}]{huensch98a}
{Huensch}, M., {Schmitt}, J.~H.~M.~M., \& {Voges}, W. 1998{\natexlab{b}},
  \aaps, 132, 155

\bibitem[{{Jansen} {et~al.}(2001){Jansen}, {Lumb}, {Altieri}, {Clavel}, {Ehle},
  {Erd}, {Gabriel}, {Guainazzi}, {Gondoin}, {Much}, {Munoz}, {Santos},
  {Schartel}, {Texier}, \& {Vacanti}}]{XMM-mission}
{Jansen}, F., {Lumb}, D., {Altieri}, B., {et~al.} 2001, \aap, 365, L1

\bibitem[{{Jordi} {et~al.}(2010){Jordi}, {Gebran}, {Carrasco}, {de Bruijne},
  {Voss}, {Fabricius}, {Knude}, {Vallenari}, {Kohley}, \& {Mora}}]{Gaia-Photo}
{Jordi}, C., {Gebran}, M., {Carrasco}, J.~M., {et~al.} 2010, \aap, 523, A48

\bibitem[{{L{\'e}pine} \& {Gaidos}(2011)}]{Lepine}
{L{\'e}pine}, S. \& {Gaidos}, E. 2011, \aj, 142, 138

\bibitem[{{Linsky} \& {Haisch}(1979)}]{linsky79}
{Linsky}, J.~L. \& {Haisch}, B.~M. 1979, \apjl, 229, L27

\bibitem[{{Pallavicini} {et~al.}(1981){Pallavicini}, {Golub}, {Rosner},
  {Vaiana}, {Ayres}, \& {Linsky}}]{pal81}
{Pallavicini}, R., {Golub}, L., {Rosner}, R., {et~al.} 1981, \apj, 248, 279

\bibitem[{{Pallavicini} {et~al.}(1990){Pallavicini}, {Tagliaferri}, \&
  {Stella}}]{pal90}
{Pallavicini}, R., {Tagliaferri}, G., \& {Stella}, L. 1990, \aap, 228, 403

\bibitem[{{Pecaut} \& {Mamajek}(2013)}]{pec13}
{Pecaut}, M.~J. \& {Mamajek}, E.~E. 2013, \apjs, 208, 9

\bibitem[{{Peres} {et~al.}(2000){Peres}, {Orlando}, {Reale}, {Rosner}, \&
  {Hudson}}]{per00}
{Peres}, G., {Orlando}, S., {Reale}, F., {Rosner}, R., \& {Hudson}, H. 2000,
  \apj, 528, 537

\bibitem[{{Pevtsov} {et~al.}(2003){Pevtsov}, {Fisher}, {Acton}, {Longcope},
  {Johns-Krull}, {Kankelborg}, \& {Metcalf}}]{pev03}
{Pevtsov}, A.~A., {Fisher}, G.~H., {Acton}, L.~W., {et~al.} 2003, \apj, 598,
  1387

\bibitem[{{Pizzolato} {et~al.}(2003){Pizzolato}, {Maggio}, {Micela},
  {Sciortino}, \& {Ventura}}]{pizzo03}
{Pizzolato}, N., {Maggio}, A., {Micela}, G., {Sciortino}, S., \& {Ventura}, P.
  2003, \aap, 397, 147

\bibitem[{{Predehl}(2017)}]{erosita}
{Predehl}, P. 2017, Astronomische Nachrichten, 338, 159

\bibitem[{{Robrade} {et~al.}(2012){Robrade}, {Schmitt}, \&
  {Favata}}]{Robrade_2012}
{Robrade}, J., {Schmitt}, J.~H.~M.~M., \& {Favata}, F. 2012, \aap, 543, A84

\bibitem[{{Saxton} {et~al.}(2008){Saxton}, {Read}, {Esquej}, {Freyberg},
  {Altieri}, \& {Bermejo}}]{sax08}
{Saxton}, R.~D., {Read}, A.~M., {Esquej}, P., {et~al.} 2008, \aap, 480, 611

\bibitem[{{Schmitt}(1997)}]{schmitt97}
{Schmitt}, J.~H.~M.~M. 1997, \aap, 318, 215

\bibitem[{{Schmitt} {et~al.}(1995{\natexlab{a}}){Schmitt}, {Fleming}, \&
  {Giampapa}}]{schmitt95}
{Schmitt}, J.~H.~M.~M., {Fleming}, T.~A., \& {Giampapa}, M.~S.
  1995{\natexlab{a}}, \apj, 450, 392

\bibitem[{{Schmitt} {et~al.}(1995{\natexlab{b}}){Schmitt}, {Fleming}, \&
  {Giampapa}}]{sch95}
{Schmitt}, J.~H.~M.~M., {Fleming}, T.~A., \& {Giampapa}, M.~S.
  1995{\natexlab{b}}, \apj, 450, 392

\bibitem[{{Schmitt} \& {Liefke}(2004)}]{schmitt04}
{Schmitt}, J.~H.~M.~M. \& {Liefke}, C. 2004, \aap, 417, 651

\bibitem[{{Skrutskie} {et~al.}(2006){Skrutskie}, {Cutri}, {Stiening},
  {Weinberg}, {Schneider}, {Carpenter}, {Beichman}, {Capps}, {Chester},
  {Elias}, {Huchra}, {Liebert}, {Lonsdale}, {Monet}, {Price}, {Seitzer},
  {Jarrett}, {Kirkpatrick}, {Gizis}, {Howard}, {Evans}, {Fowler}, {Fullmer},
  {Hurt}, {Light}, {Kopan}, {Marsh}, {McCallon}, {Tam}, {Van Dyk}, \&
  {Wheelock}}]{2MASS}
{Skrutskie}, M.~F., {Cutri}, R.~M., {Stiening}, R., {et~al.} 2006, \aj, 131,
  1163

\bibitem[{{Stelzer} {et~al.}(2006){Stelzer}, {Schmitt}, {Micela}, \&
  {Liefke}}]{stelzer06}
{Stelzer}, B., {Schmitt}, J.~H.~M.~M., {Micela}, G., \& {Liefke}, C. 2006,
  \aap, 460, L35

\bibitem[{{Stocke} {et~al.}(1991){Stocke}, {Morris}, {Gioia}, {Maccacaro},
  {Schild}, {Wolter}, {Fleming}, \& {Henry}}]{sto91}
{Stocke}, J.~T., {Morris}, S.~L., {Gioia}, I.~M., {et~al.} 1991, \apjs, 76, 813

\bibitem[{{Vaiana} {et~al.}(1981){Vaiana}, {Cassinelli}, {Fabbiano},
  {Giacconi}, {Golub}, {Gorenstein}, {Haisch}, {Harnden}, {Johnson}, {Linsky},
  {Maxson}, {Mewe}, {Rosner}, {Seward}, {Topka}, \& {Zwaan}}]{vai81}
{Vaiana}, G.~S., {Cassinelli}, J.~P., {Fabbiano}, G., {et~al.} 1981, \apj, 245,
  163

\bibitem[{{Vilhu}(1984)}]{vil84}
{Vilhu}, O. 1984, \aap, 133, 117

\bibitem[{{Walter} {et~al.}(1978){Walter}, {Charles}, \& {Bowyer}}]{wal78}
{Walter}, F., {Charles}, P., \& {Bowyer}, S. 1978, \apjl, 225, L119

\bibitem[{{Weisskopf} {et~al.}(2002){Weisskopf}, {Brinkman}, {Canizares},
  {Garmire}, {Murray}, \& {Van Speybroeck}}]{Chandra-mission2}
{Weisskopf}, M.~C., {Brinkman}, B., {Canizares}, C., {et~al.} 2002, \pasp, 114,
  1

\bibitem[{{Weisskopf} {et~al.}(2000){Weisskopf}, {Tananbaum}, {Van Speybroeck},
  \& {O'Dell}}]{Chandra-mission1}
{Weisskopf}, M.~C., {Tananbaum}, H.~D., {Van Speybroeck}, L.~P., \& {O'Dell},
  S.~L. 2000, in \procspie, Vol. 4012, X-Ray Optics, Instruments, and Missions
  III, ed. J.~E. {Truemper} \& B.~{Aschenbach}, 2--16

\bibitem[{{Wenger} {et~al.}(2000){Wenger}, {Ochsenbein}, {Egret}, {Dubois},
  {Bonnarel}, {Borde}, {Genova}, {Jasniewicz}, {Lalo{\"e}}, {Lesteven}, \&
  {Monier}}]{SIMBAD-database}
{Wenger}, M., {Ochsenbein}, F., {Egret}, D., {et~al.} 2000, \aaps, 143, 9

\bibitem[{{Worthey} \& {Lee}(2011)}]{color-table}
{Worthey}, G. \& {Lee}, H.-c. 2011, \apjs, 193, 1

\end{thebibliography}

\appendix
\section{Flowchart of the matching and cleaning procedure}
\label{sec: flowchart}
\begin{figure}[H]
\centering
  \includegraphics[width=\hsize]{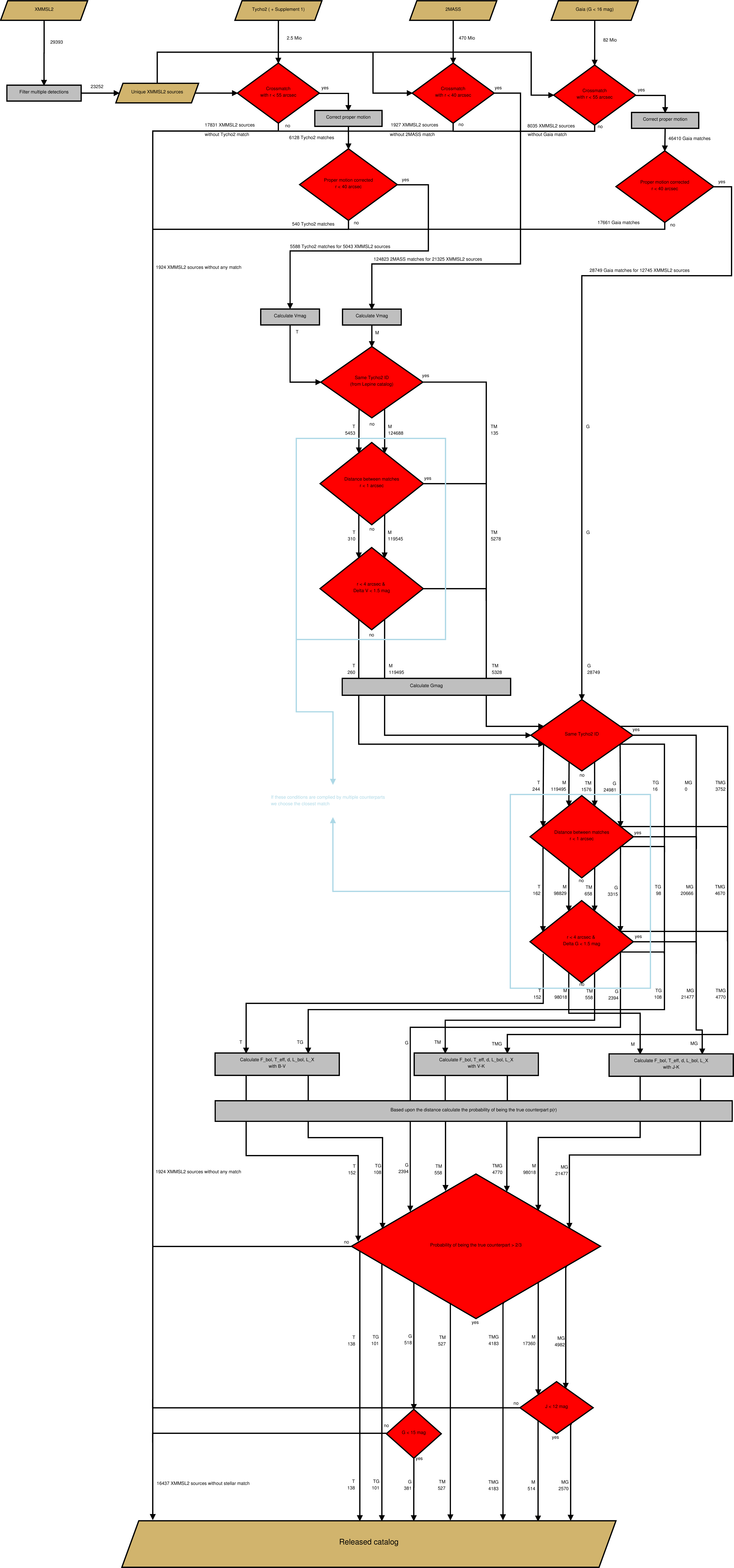}
  \caption{Flowchart of the matching procedure}
  \label{fig: flowchart}
\end{figure}

\begin{figure}[H]
\centering
  \includegraphics[height=\textheight]{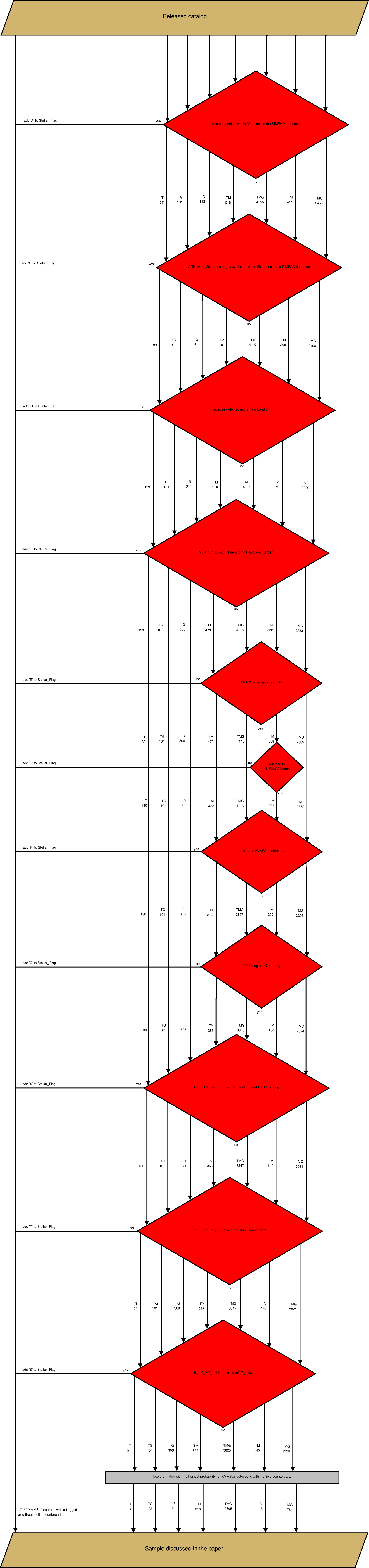}
  \caption{Flowchart of the cleaning procedure}
  \label{fig: flowchart cleaning procedure}
\end{figure}

\section{Column description}
\label{sec: column description}
We adopt all columns of the XMMSL2 catalog, described at \url{https://www.cosmos.esa.int/web/xmm-newton/xmmsl2-ug}, and extend further 57 columns defining our stellar counterpart. In the following we describe these columns. For measurements given in multiple catalogs, we define an order of priority from which catalog the values are adopted. 

\subsubsection*{F\_X\_XMMSL}
We convert the count rates given in XMMSL2 catalog into a 0.1\,--\,2.4~keV X-ray flux (cf. Sect~\ref{sec: XMMSL2 catalog}). We use the count rate of the total band if available, otherwise we adopt the count rate of the soft band or the hard band.\\*
Units: $10^{-12}\;\mathrm{erg\,cm^{-2}\,s^{-1}}$\\*
Order of priority: total band, soft band, hard band\\

\subsubsection*{num\_detections}
Number of detections of the XMMSL2 source

\subsubsection*{num\_match}
Number of stellar counterparts of the XMMSL2 source

\subsubsection*{Priority}
For XMMSL2 sources with multiple counterparts, the counterparts are sorted by the matching probability ''prob\_r''. This column gives the ranking of the counterparts so that the identification with the highest matching probability is set to ''Priority=1''. 

\subsubsection*{Catalog}
Catalog of the stellar counterpart \\
G: Gaia catalog\\
T: Tycho2 catalog\\
M: 2MASS catalog\\
L: Lepine catalog\\
B: BrightStar catalog\\
For counterparts given in multiple catalogs, multiple abbreviations are given, e.g., ''TMG''.

\subsubsection*{r}
The distance $r$ between the position given in the XMMSL2 catalog and the proper motion corrected position of the stellar counterpart\\*
Units: arcsec\\*
Order of priority: Gaia catalog, Tycho2 catalog, Lepine catalog, BrightStar catalog, 2MASS catalog

\subsubsection*{prob\_r}
Matching probability as defined in Eq.~\ref{equ: probability function} of the stellar counterpart \\*
Units: \%\\*
Order of priority: Lepine catalog, BrightStar catalog, Tycho2 catalog, 2MASS catalog, Gaia catalog

\subsubsection*{Gaia\_ID}
Identifier of the Gaia catalog 

\subsubsection*{Tycho2\_ID}
Identifier of the Tycho2 catalog 

\subsubsection*{HR}
Identifier of the BrightStar catalog

\subsubsection*{PM}
Identifier of the SUPERBLINK catalog given in the Lepine catalog

\subsubsection*{2MASS\_ID}
Identifier of the 2MASS catalog

\subsubsection*{Match\_RA, Match\_DEC}
Proper motion corrected position of the stellar counterpart \\* 
Units: degrees\\*
Order of priority: Gaia catalog, Tycho2 catalog, Lepine catalog, BrightStar catalog, 2MASS catalog

\subsubsection*{e\_RA, e\_DEC}
Statistical error on the position of the stellar counterparts if it is available\\*
Units: mas

\subsubsection*{pmRA, pmDEC, e\_pmRA, e\_pmDEC}
Proper motion and their statistical errors of the stellar counterpart \\*
Units: $\mathrm{mas\,yr^{-1}}$\\*
Order of priority: Tycho2 catalog, Lepine catalog, BrightStar catalog 

\subsubsection*{BTmag, e\_BTmag, VTmag, e\_VTmag}
Magnitude $B_T$ and $V_T$ and their statistical errors adopted from the Tycho2 catalog.\\*
Units: mag

\subsubsection*{Bmag}
The B magnitude is given in the Lepine catalog, but originate from the USNO catalog. It can also be estimated from the color $B-V$ given in the BrightStar catalog or from the Tycho2 colors \citep{hip97}.\\*
Units: mag\\* 
Order of priority: Tycho2 catalog, BrightStar catalog, Lepine catalog

\subsubsection*{Vmag}
The V magnitude is directly measured in the BrightStar catalog. In the Lepine catalog the V magnitude is also given, but it is estimated from magnitudes of other photometric bands. We can estimate the V magnitude from the Tycho2 colors \citep{hip97} or extrapolate it from the color $J-K$ of the 2MASS catalog. \\*
Units: mag\\* 
Order of priority: Tycho2 catalog, BrightStar catalog, Lepine catalog, 2MASS catalog

\subsubsection*{Gmag}
G magnitude adopted from the Gaia catalog. \\*
Units: mag

\subsubsection*{Jmag, e\_Jmag, Hmag, e\_Hmag, Kmag, e\_Kmag}
Magnitudes in the J, H and K bands and the statistical errors adopted from the 2MASS catalog.\\*
Units: mag

\subsubsection*{2MASS\_ph\_qual}
Quality of the 2MASS magnitude measurement adopted from the 2MASS catalog

\subsubsection*{2MASS\_rd\_flg}
This flag describes for every 2MASS photometric band, which method has been applied to determine the magnitude. If the flag contains a ''0'', the source is not detected in that band. We flag these sources if they do not have an entry in another catalog.  

\subsubsection*{2MASS\_bl\_flg}
Number of sources for which the 2MASS magnitude is estimated at the same time

\subsubsection*{2MASS\_cc\_flg}
This flag indicates if the magnitude or the position of a 2MASS source is contaminated by a nearby source. 

\subsubsection*{2MASS\_gal\_contam}
This flag indicates if a 2MASS point source lies within the boundaries of an extended 2MASS source.

\subsubsection*{2MASS\_ext\_key}
If a 2MASS point source is associated with an extended 2MASS source, this column gives the identifier of the extended source.

\subsubsection*{Parallax, e\_Parallax}
Trigonometric parallax and their statistical errors if it is known \\*
Units: arcsec\\* 
Order of priority: Gaia catalog, Hipparcos catalog, Lepine catalog, BrightStar catalog \\

\subsubsection*{CCDM}
Identifier of the component of multiple star system adopted from the Tycho2 or the BrightStar catalog\\*
Order of priority: Tycho2 catalog, BrightStar catalog\\

\subsubsection*{SpType}
Spectral type adopted from the BrightStar or Lepine catalog\\*
Order of priority: BrightStar catalog, Lepine catalog\\

\subsubsection*{m\_bol} 
We estimate the bolometric magnitude using Table~3 of \citet{color-table}. For the calculation we use different colors in the following order of priority.\\*
Units: mag\\* 
Order of priority: $V-K$, $B-V$, $J-K$\\

\subsubsection*{F\_bol}
Given the bolometric magnitude we estimate the bolometric flux.\\*
Units: $10^{-12}\;\mathrm{erg\,cm^{-2}\,s^{-1}}$

\subsubsection*{T\_eff}
Effective temperature given in Table~3 of \citet{color-table}\\*
Units: K\\* 
Order of priority: $V-K$, $B-V$, $J-K$\\

%\subsubsection*{d}
%The distance of the source can be determined with the trigonometric parallax if it is known, otherwise we estimate the photometric distance adopting the data of Pecaut \& Mamajek (2013).\\*
%Order of priority: parallax, $V-K$, $B-V$, $J-K$\\

\subsubsection*{abs\_M\_V, abs\_M\_bol}
Given the trigonometric parallax we estimate the absolute magnitude.\\*
Units: mag

\subsubsection*{L\_bol, L\_X}
Given the trigonometric parallax we estimate the luminosity.\\*
Units: $10^{30}\;\mathrm{erg\,s^{-1}}$

\subsubsection*{Stellar\_Flag}
\begin{table}[H]
\footnotesize
\begin{tabular}{p{0.2cm}p{3.0cm}p{4.0cm}p{0.4cm}}
\hline\hline
Sy. & Description & Definition & No. \\
\hline
A & Accretor warning & Accreting objects within $30$~arcsec in the SIMBAD database & 111 \\
G & Extragalactic warning & AGN within $30$~arcsec or a galaxy cluster within $60$~arcsec in the SIMBAD database & 116 \\
O & Optical loading & ''VER\_OPTLOAD'' is true and no RASS counterpart & 59 \\
E & Extended source & ''ext\_key'' $\neq 0$ in the 2MASS catalog & 21 \\
D & Missing 2MASS detection & 2MASS counterpart only and ''rd\_flg'' = 0 in one band & 22 \\
P & Erroneous 2MASS photometry & ''ph\_qual'' = X, U, F or E in one band & 467 \\
C & Extreme color & $J-K<-0.25$~mag or $J-K>1$~mag & 311 \\
H & Hard band detection only & No values given in ''RATE\_B8'' and ''RATE\_B6'' & 3 \\
X & Persistent high X-ray activity & $\log(F_\mathrm{X}/F_\mathrm{bol} > -2.2$ in the XMMSL2 and RASS catalog & 61 \\
T & Transient X-ray source & $\log(F_\mathrm{X,XMMSL2}/F_\mathrm{bol} > -1.5$ and no RASS counterpart & 22 \\
S & High X-ray activity in the slew & Fractional contribution of the X-ray flux to the bolometric flux in the slew higher than usual for a source with the specific effective temperature (cf. Fig.~\ref{fig: F_x-F_bol}) & 123 \\
\end{tabular}
\end{table}
\subsubsection*{RASS\_ID}
Identifier of the closest RASS counterpart up to a distance of $60$~arcsec

\subsubsection*{RASS\_r}
Distance between the XMMSL2 source and the closest RASS counterpart\\*
Units: arcsec

\subsubsection*{RASS\_F\_X}
RASS flux of the closest RASS counterpart adopting the conversion factor defined by \citet{sch95}\\*
Units: $10^{-12}\;\mathrm{erg\,cm^{-2}\,s^{-1}}$

\subsubsection*{Simbad\_ID}
Identifier of the closest SIMBAD counterpart up to a distance of $30$~arcsec

\subsubsection*{Simbad\_r}
Distance between the XMMSL2 source and the closest SIMBAD counterpart\\*
Units: arcsec

\subsubsection*{Simbad\_otype}
Classification of the closest SIMBAD counterpart

\end{document}